\documentclass[twocolumn,aps,showpacs,prb,amsmath,amssymb,floatfix,superscriptaddress]{revtex4-2}

\usepackage{amssymb,color}

\usepackage{bm} 
\usepackage{color}
\usepackage{graphicx}
\usepackage{physics}
\usepackage{amsthm}
\usepackage{amsmath}
\usepackage{amssymb}
\usepackage{enumerate}
\usepackage{placeins}
\usepackage{booktabs}
\usepackage{dsfont}
\usepackage{yfonts}
\usepackage{hyperref} 

\DeclareSymbolFont{bbold}{U}{bbold}{m}{n}
\DeclareSymbolFontAlphabet{\mathbbold}{bbold}

\newcommand{\cord}[1]{{\sin\left(\frac{\pi}{L}\left(#1\right)\right)}}  
\newcommand{\eps}{{\varepsilon}}  
  
\newcommand{\ex}[1]{\langle #1 \rangle}

\newcommand{\ren}{R\'{e}nyi~}
\newcommand{\Lb}{L}

\begin{document}

\title{One-particle entanglement for one dimensional spinless fermions after an interaction quantum quench}

\author{Matthias Thamm}
\affiliation{Institut f\"{u}r Theoretische Physik, Universit\"{a}t Leipzig,  Br\"{u}derstrasse 16, 04103 Leipzig, Germany}

\author{Harini Radhakrishnan}
\affiliation{Department of Physics and Astronomy, University of Tennessee, Knoxville, TN 37996, USA}
\affiliation{Institute for Advanced Materials and Manufacturing, University of Tennessee, Knoxville, Tennessee 37996, USA\looseness=-1}

\author{Hatem Barghathi}
\affiliation{Department of Physics and Astronomy, University of Tennessee, Knoxville, TN 37996, USA}
\affiliation{Institute for Advanced Materials and Manufacturing, University of Tennessee, Knoxville, Tennessee 37996, USA\looseness=-1}

\author{Bernd Rosenow}
\affiliation{Institut f\"{u}r Theoretische Physik, Universit\"{a}t Leipzig,  Br\"{u}derstrasse 16, 04103 Leipzig, Germany}

\author{Adrian Del Maestro}
\affiliation{Department of Physics and Astronomy, University of Tennessee, Knoxville, TN 37996, USA}
\affiliation{Institute for Advanced Materials and Manufacturing, University of Tennessee, Knoxville, Tennessee 37996, USA\looseness=-1}
\affiliation{Min H. Kao Department of Electrical Engineering and Computer Science, University of Tennessee, Knoxville, TN 37996, USA}

\date{\today}

\begin{abstract}
Particle entanglement provides information on quantum correlations in systems of indistinguishable particles. Here, we study the one particle entanglement entropy for an integrable model of spinless, interacting fermions both at equilibrium and after an interaction quantum quench. Using both large scale exact diagonalization and time dependent density matrix renormalization group calculations, we numerically compute the one body reduced density matrix for the $J$-$V$ model, as well as its post-quench dynamics.  We include an analysis of the fermionic momentum distribution, showcasing its time evolution after a quantum quench. Our numerical results, extrapolated to the thermodynamic limit, can be compared with field theoretic bosonization in the Tomonaga-Luttinger liquid regime. Excellent agreement is obtained using an interaction cutoff that can be determined uniquely in the ground state. 
\end{abstract}
 
\maketitle

\section{Introduction}

If a quantum system is in a pure state after a sudden change to the system --  a quantum quench \cite{Calabrese:2006kr} --  the growth of entanglement entropy under a spatial bipartition plays
the role of thermal entropy \cite{Calabrese:2005oi,Alba:2017ph} in describing how expectation values of local observables can be computed from a statistical ensemble
\cite{Srednicki:1994xg, Rigol:2008sl,Polkovnikov:2011gg,DAlessio:2016rr}. 
 For systems of $N$ indistinguishable particles, a bipartition can also be made in terms of subgroups of $n$ and $N-n$ particles \cite{Zanardi:2002xc,Shi:2003ro,Zozulya:2008bg,Haque:2009zi}. 
This particle entanglement entropy provides complementary information as compared to the spatial mode entanglement and is sensitive to both interactions and particle statistics at leading order \cite{Haque:2007mu,Zozulya:2007fe,Santachiara:2007it,Katsura:2007ty,Herdman:2014jy,Herdman:2014vd,Herdman:2015xa,Barghathi:2017ab,Rammelmueller:2017al,Iemini:2015ze,Ferreira:2022qq,Pu:2022pe} and possibly to many body localization \cite{Lin2018,Hopjan:2021tl}. An equivalence was recently demonstrated between the increase of entropy densities under spatial and particle bipartitions in the asymptotic steady-state regime after a quantum quench in a system of interacting one dimensional fermions \cite{DelMaestro:2021ja} in the limit $n,N \to \infty$; $n/N \sim \text{const}$.   

Numerical results also indicate that the particle entanglement entropy density is a decreasing function of the order $n$ of the reduced density matrix, which can be understood in terms of higher-order correlations acting as a constraint on the available particle configurations. 
For an integrable model, it is even possible to obtain the entanglement entropy density after an interaction quench from knowledge of only the diagonal components of the $n=2$ density matrix \cite{DelMaestro:2022mnp}. These results accentuate the potential of particle entanglement entropy as an alternative to the usual spatial entanglement in understanding non-classical correlations in non-equilibrium quantum dynamics.

Thus it is natural to explore the entanglement entropy associated with low order density matrices. The idea of expanding the entropy density as a series in irreducible correlations between groups of $n$ particles is explicit in classical non-equilibrium mechanics \cite{Kirkwood42,Green52}. However, in general, density matrices are very challenging to compute, but in one dimension (1d), even after a quantum quench, bosonization gives access to low-order reduced density matrices  as correlation functions of bosonic exponentials \cite{Cazalilla:2006zw,Uhrig:2009zd,Iucci:2009lg,Dora:2011or}.  
Here,  we study the properties of the $n=1$ reduced density matrix (RDM), both in equilibrium and after a quantum quench, with a focus on the von Neumann and \ren entropies. This represents the first step in the systematic expansion in terms of multi-particle correlations discussed above. For $n=1$, the 1-RDM is proportional to the familiar equal time Green function which captures the momentum distribution, and is experimentally accessible in a wide variety of scenarios (e.g.\@ via interference \cite{Polkovnikov:2006js,Gritsev:2006tn} or Raman scattering \cite{Dao:2007pl} in trapped low dimensional ultracold gasses, or through the spectral function in angle resolved photoemission spectroscopy \cite{Damascelli:2004db}).  Bounds have been proven on the spectrum of  1-RDMS, and there are conjectures for the spectrum of the 2-RDM \cite{Carlen:2016fv,Lemm:2017ha}.

We perform exact computations of the 1-RDM for an interacting lattice model of spinless fermions in one dimension, the $J$-$V$ model. This model, which can be exactly solved by mapping to the $XXZ$ spin chain \cite{DesCloizeaux:1966,Yang:1966yc} at fixed magnetization, has proven to be a fruitful playground for studying quasi-thermalization and the dynamics of correlation functions and spatial entanglement after a quantum quench \cite{Manmana:2007tw,Manmana2009,Rigol2009,Foster2011,Coira2013}.  Here, we are interested in particle entanglement entropy in this system, and apply large scale exact diagonalization (up to $N=19$ particles on $L=38$ sites at half filling), both in equilibrium and after an interaction quantum quench. Both the transient dynamics and asymptotic steady state after a quantum quench are analyzed by performing unitary time evolution starting from an initial state of free fermions to long times.  These results are extended to even larger system sizes while preserving periodic boundary conditions using GPU accelerated time-dependent density matrix renormalization group (tDMRG) calculations allowing us to study systems up to $L=102$ sites.  Here, the presence of periodic boundary conditions is important, allowing for the computation of the momentum distribution directly from the eigenvalues of the 1-RDM, maintaining translational invariance and ensuring accuracy of measured quantities at small momenta.

We consider a wide range of attractive and repulsive interactions spanning a continuous and discrete quantum phase transition in the model. For a quantum quench to a state with strong interactions, outside the quantum liquid regime, we find that both the transient and long time momentum distribution can develop non-monotonic behavior as a function of momentum $q$ -- a signature of strong spatial correlations and particle localization in the ground state. 
The large system sizes studied allow us to perform reliable finite size scaling to the thermodynamic limit where a comparison can be made with continuum field theory calculations. 

Bosonization is routinely used to compute universal quantities, and here we use it to study the 1-RDM whose short-distance behavior reflects the dynamics of high energy excitations. This is accomplished by the introduction of an interaction cutoff (different from the often used UV lattice cutoff), which is needed due to the short-range nature of interactions in the $J$-$V$ model under study. This cutoff is unambiguously determined from our equilibrium numerics and applied to make microscopic predictions after the quench via bosonization. Good agreement is found across the phase diagram for the 1-particle von Neumann and R\'{e}nyi entanglement entropies highlighting the utility of continuum field theory to describe both short and long time dynamics.  

\begin{figure}[t!]
			\centering
			\includegraphics[width=8.6cm]{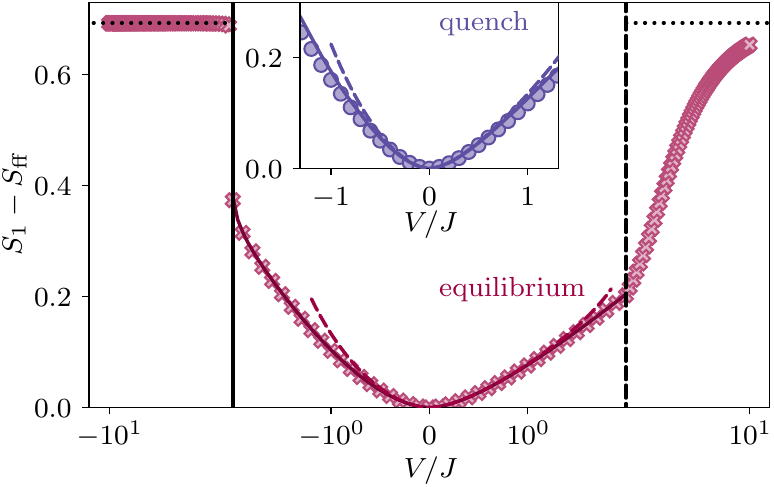}  
			\caption{\label{Fig:introd001}Interaction dependence of one particle von Neumann entanglement entropy $S_1$ obtained numerically from the $J$-$V$ model and from an effective low energy Luttinger liquid calculation (dashed lines) with a fixed interaction cutoff. Here, $S_{\rm ff}$ is the entropy for free fermions.  The main panel depicts the equilibrium ground state entropy with numerical data from DMRG for a system of $\Lb=102$ lattice sites at half filling (crosses). The solid line represents finite size scaling of numerical data to the thermodynamic limit.  The excellent agreement with the finite size DMRG data shows that the system with $N=51$ fermions is large enough to describe the thermodynamic limit accurately in the whole LL phase. 
			The inset depicts finite size exact diagonalization results for $N=12$ fermions on $L=24$ sites after an interaction quantum quench (circles) in the asymptotic steady state. The solid line shows the thermodynamic limit of the numerical data obtained from finite size scaling the time averaged one particle entropy (circles) after the interaction quench.  The dashed line is the result of non-equilibrium bosonization using the same value of the interaction cutoff as in the main panel.}
\end{figure}

In the following, we briefly describe the main results and contributions of this work.  We study $N$ one dimensional spinless fermions on a lattice of $L$ sites with hopping $J$ and nearest neighbor interaction $V$ (see microscopic Hamiltonian in Eq.~\ref{Eq:eq_JVHam} for details). For $\abs{V/J}<2$, the low energy sector is a Luttinger liquid, while the system undergoes a continuous quantum phase transition to an insulating solid phase at $V/J =2$.  For attractive interactions, there is a discontinuous transition to a phase separated clustered solid at $V/J = -2$. This phase diagram is reflected in Figure~\ref{Fig:introd001} which shows the von Neumann entanglement entropy $S_1$ computed from the spectrum of the 1-RDM, $\rho_1(q)$: 
\begin{equation}
       S_1 =  -\frac{N}{2k_F}\int dq\; \rho_1(q)\ln \rho_1(q)   \ ,
\end{equation}
where $k_{\rm F}$ is the Fermi momentum at half-filling. 
Here we have subtracted off the 1-particle entanglement of free fermions: $S_{\rm ff} = \ln N$ to highlight the role of interactions. At $V/J =2$, there is a change of slope in the entanglement as the system enters the solid phase via a second order transition, and $S_1-S_{\rm ff}$ asymptotically approaches $\ln 2$ (dotted line) for $V/J \gg 2$ reflecting the two-fold degeneracy of the charge density wave ground state.   Moving across $V/J = -2$, the entanglement entropy echoes the first-order transition by a sudden jump in $S_1-S_{\rm ff}$ to $\approx\ln 2$ (dotted line). Here, the large entanglement entropy is due to the translational symmetry of the clustered $N$ fermions state representing the solid phase. 
In the Luttinger liquid phase, we show the bosonization result for the entanglement as a dashed red line, for a fixed value of the interaction cutoff. The deviation for strong negative interactions reflects the divergence $K \to \infty$ of the Luttinger parameter when approaching the first order phase transition at $V/J = -2$. The solid red line represents the extrapolation to the thermodynamic limit of the numerical exact diagonalization and DMRG data, highlighting the reliable nature of our finite size scaling procedure. The inset shows the $t\to\infty$ asymptotic limit of the 1-particle entanglement entropy after the $J$-$V$ model is quenched from non-interacting fermions to a final interaction strength $V$ at $t=0$ for a finite size system of $N=12$ fermions on $L=24$ sites. Here, the dashed line is computed via non-equilibrium bosonization using the same cutoff as in the equilibrium case. The solid purple line again shows the extrapolation to the thermodynamic limit. 
A comparison with the main panel shows the growth of entanglement after the quantum quench in this quantum liquid regime. 

\begin{figure}[ t! ]
			\centering
			\includegraphics[width=8.6cm]{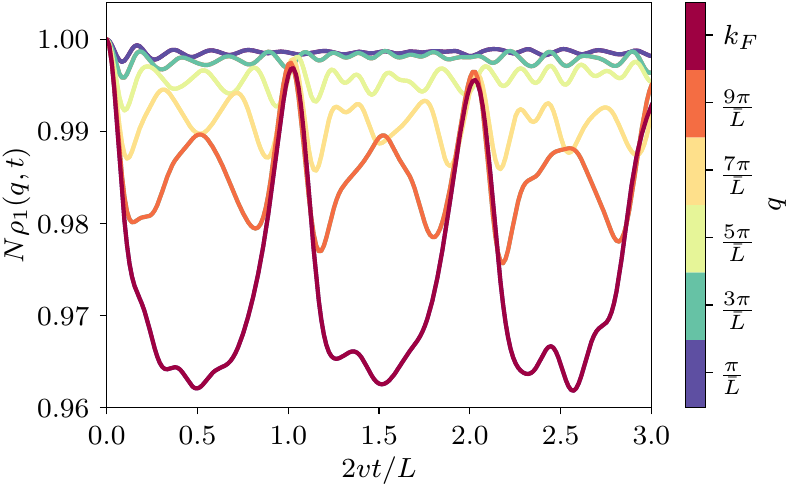}  
			\caption{\label{Fig:introd002}Dependence of eigenvalues of the one body density matrix $\rho_1$ on the rescaled waiting time $2vt/L$ after the quantum quench computed via exact diagonalization for a final interaction strength $V/J = -0.5$. We show a quarter of the spectrum with $0<q<k_F$ for a system of $N=12$ fermions on $\Lb=24$ lattice sites. The largest contributions to the one particle entanglement entropy come from the eigenvalues close to the Fermi levels which show the largest oscillation amplitude and the recurrence time $L/(2v)$ that appears in the entropy.}
    \end{figure}

In translationally invariant systems, the 1-RDM depends only on the difference between the two spatial coordinates, and can hence be diagonalized by a Fourier transform. The resulting $\rho(q,t)$ obtained from exact diagonalization is displayed in Fig.~\ref{Fig:introd002} as a function of $2 v t/L$, where $t$ is the waiting time after the quench,  $v$ is the renormalized velocity of low energy excitations, and $L$ is the system size. 
Quasi-periodic oscillations, due to the presence of multiple velocity scales, whose amplitude increases with momentum $q$ are observed. 

\begin{figure}[t!]
\centering
\includegraphics[width=8.6cm]{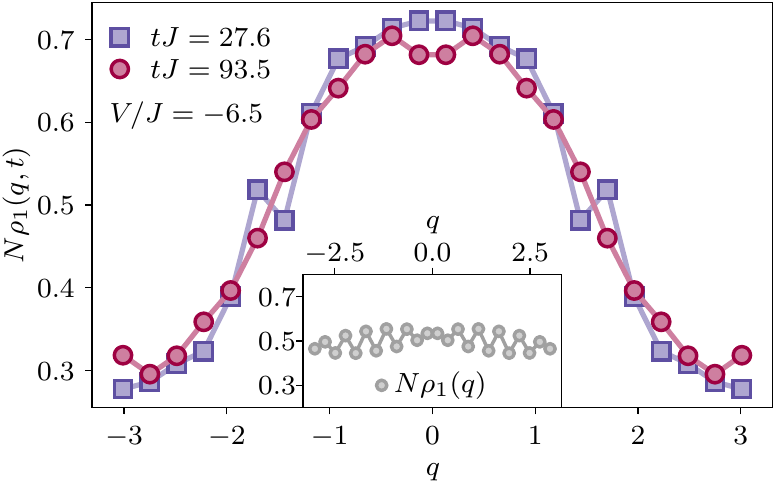}  
\caption{\label{Fig:sp_obdmQtd} Momentum distribution function at two different times, after a quantum quench at $t=0$ to a final interaction strength $V/J = -6.5$ deep in the phase separated clustered solid. The system consists of $N = 12$ fermions on a ring of $L = 24$ sites.  The inset shows the equilibrium ground state distribution function which demonstrates pronounced oscillations due to the existence of a short momentum scale resulting from large clusters of $N$ fermions. Lines are a guide to the eye.}
\end{figure}

For a strong interaction quench from non-interacting fermions to deep inside the phase separated cluster solid, the $q$-dependence of the distribution function at fixed waiting times can develop a non-monotonicity as seen in Figure~\ref{Fig:sp_obdmQtd}. For times still in the transient range, this can occur near the Fermi momentum, whereas, at long waiting times, it appears even at small $q$.  
A detailed study of the interaction and time dependence of this quantity is discussed in Section \ref{sec:time_dependence_rho}.
%%%%%%%%%%%%%%%%
%%%%%%%%%%%%%%%%

The main contributions of this work are i) providing a definitive picture of the 1-RDM and entanglement in a one-dimensional integrable model both in equilibrium and after a quantum quench; and ii) utilizing a self-consistent procedure to regularize field theory computations to make predictions about the post-quench density matrix and entanglement entropy. 

The remainder of the manuscript is organized as follows.  In Section II we introduce the microscopic lattice model under study and describe its phase diagram in detail.  We then bosonize its low energy sector in Section III and derive an expression for the momentum distribution in equilibrium.  This field theory calculation is then compared against exact diagonalization and DMRG results in Section IV. Section V explores the 1-particle entanglement entropy after an interaction quantum quench, again comparing field theory with numerical results. The explicit post-quench waiting time dependence of the 1-RDM is investigated in Section VI before we provide some final concluding remarks and possible future research directions in Section VII.

\section{Model and Phase Diagram}
\label{sec:model}

    % J-V model Hamiltonian
    We study a system of $N$ spinless fermions on a one-dimensional lattice with $\Lb$ sites at half-filling $\Lb=2N$ described by the $J$-$V$ Hamiltonian 
    \begin{align}
        H &= -J \sum_{i=1}^{\Lb} (c_{i+1}^\dagger c_i^{\phantom{\dagger}} + c_i^\dagger c_{i+1}^{\phantom{\dagger}}) + V \sum_{i=1}^{\Lb} n_i n_{i+1} \ . \label{Eq:eq_JVHam}
    \end{align}
    Here $J$ is the hopping amplitude, $V$ is the nearest neighbor interaction, $c_i^\dagger$ creates a fermion at site $i$, and $n_i=c_i^\dagger c_i$ is the occupation number operator for site $i$. In the case of even number of particles $N$ we use anti-periodic boundary conditions and for odd $N$ we use periodic boundary conditions, which ensures that the ground state is always non-degenerate.
    
    % Phases of J-V model
    Depending on the strength of the interaction parameter $V/J$, the system is in one of three phases, where the exact phase boundaries are known from a mapping to a spin-$1/2$ XXZ model \cite{DesCloizeaux.1966,Yang.1966}:
    \begin{enumerate}[i)]
        \item For $V/J<-2$ the system is a phase separated solid, where the strong attractive interactions favor clustering of fermions such that the ground state for $V/J\rightarrow-\infty$ becomes  
        \begin{align}
            \ket{\Psi_{V/J\rightarrow-\infty}} 
            &= \frac{1}{\sqrt{L}} \sum_{n=1}^L T^n \ket{11\cdots1100\cdots0} \ ,
            \label{Eq:eq_JV_psiPinf} 
        \end{align}
        where $\ket{11\cdots1100\cdots0}$ is the state for which the first $N$ sites are occupied by a fermion and the remaining $N$ sites are empty.
        Here, $T$ is the translation operator that shifts each fermion one site to the right, \emph{e.g}.~$T\ket{011001}=\ket{101100}$. 
        \item For the other strongly interacting case $V/J>2$, the system is in the charge density wave phase where strong repulsion results in a ground state with maximal separations between the fermions. For a lattice at half filling, the ground state in the limit $V/J\rightarrow\infty$  becomes 
        \begin{align}
        \ket{\Psi_{V/J\rightarrow+\infty}} 
            &= \frac{1}{\sqrt{2}} \left( \ket{10101\cdots} + \ket{01010\cdots} \right) \ .
            \label{Eq:eq_JV_psiMinf}
        \end{align}
        \item In the intermediate region, $-2<V/J<2$, the system is in the Tomonaga-Luttinger liquid (LL) phase, where the relatively weak interactions allow the description with an effective low-energy theory.
    \end{enumerate} 
    
\section{1-Particle Reduced Density Matrix in a Luttinger Liquid} 

    In this paper, we study the one-particle entanglement entropy from the $J$-$V$ model in the LL phase (iii). In the following, we measure lengths in units of the lattice constant. We start with deriving an analytical result for the one body density matrix $\rho_1(x,0)$ 
    for the corresponding LL model of length $L$. From the LL 1-RDM, we compute the  one-particle entanglement entropy, and then compare with numerical results obtained for the $J$-$V$ model. 
    In this phase with intermediate  interaction strength, observables are dominated by low
    energy excitations in the form of density fluctuations around a static
    average density background. Such fluctuations of the density are bosonic in
    nature, which allows us to describe the low energy physics with an effective
    Hamiltonian \cite{Iucci:2009lg} in bosonization notation after linearizing the dispersion around the Fermi points
    \begin{equation}
    \begin{aligned}
       H &= \sum_{q \neq 0} \left[\omega_0(q)+m(q)\right] b_q^\dagger b_q \\
         & \ \ + \frac{1}{2} \sum_{q \neq 0} g_2(q) \left(b_qb_{-q}+b_q^\dagger b_{-q}^\dagger\right)\ , 
    \end{aligned}
    \label{Eq:eq_LL_Hbos} 
\end{equation}
    where  $b_q^\dagger$ ($b_q$) are bosonic creation (annihilation) operators with $[b_q,b_{q'}^\dagger]=\delta_{q,q'}$, $\omega_0(q) = v_F |q|$, and we work in units where $\hbar=1$. The sum is taken over discrete momenta $q_n = n 2 \pi/L$ with $n \in \mathbb{Z} \setminus \{0\}$. 
 The nearest neighbor coupling in the lattice model has a finite interaction range, which we take into account by 
  assuming that $g_2(q)$ and $m(q)$ vanish for momenta  $q \eps \gg 1$ larger than an interaction cutoff $\eps$. 
\footnote{for the detailed implementation of the cutoff procedure see Eq.~(\ref{Eq:eq_LL_gammaeq})}.  By comparing the final bosonization results  with numerical simulations of the $J$-$V$ chain
 the interaction cutoff can be unambiguously determined. 
    The Hamiltonian Eq.~\eqref{Eq:eq_LL_Hbos} is quadratic in the boson operators and can therefore be solved analytically. In order to compute the one body density matrix, we use  refermionization to express the fermionic field operators $\psi_\alpha(x)$ in terms of bosonic fields as
    \begin{align}
        \psi_\alpha(x)&=\frac{\chi_\alpha}{\sqrt{2\pi \eta}}e^{\imath (\varphi_{0,\alpha} +\alpha \frac{2\pi x}{L} N_\alpha)  } e^{-\imath\phi_\alpha(x)} \label{Eq:eq_LL_psi}\\ 
        \phi_\alpha(x) &= - \sum_{q>0} \sqrt{\frac{2\pi}{qL}} e^{-q \eta/2}  
        \Big[ e^{\imath \alpha q x} b_{\alpha q}   + e^{- \imath \alpha q x }b_{\alpha q}^\dagger \Big] \ , \label{Eq:eq_LL_phi}
    \end{align}
    where $\alpha=(-)1$ indicates right (left) moving fermions,  $\chi_\alpha = e^{\alpha \imath \frac{\pi}{2} N_{- \alpha}}$ are Klein factors with $\chi_\alpha^\dagger\chi_\alpha=1$, $\eta$ is a short distance cutoff measured in units of the lattice spacing (not to be confused with the interaction cutoff $\eps$), and $\phi_\alpha(x)$ are Hermitian operators \cite{Giamarchi.2010, Eggert.2009}.  Here, $N_\alpha$ is the particle number operator, and $\varphi_{0,\alpha}$, $N_\alpha$  are zero mode operators satisfying the commutation relation  $[N_\alpha,\varphi_{0,\alpha}] = i$. 
    The one-body density matrix can be obtained from the one point correlation functions for left and right movers in terms of the fermion operators Eq.~\eqref{Eq:eq_LL_psi} via 
    \begin{align}
        \rho_1(x,0)&= \frac{1}{N} \qty[ e^{-\imath k_F x} C_+(x,0) + e^{\imath k_F x}C_-(x,0) ] 
        \label{Eq:eq_LL_obdm_1}\\
        C_\alpha(x,0) &= \ex{\psi_\alpha^\dagger(x)\psi_\alpha(0)} \label{Eq:eq_LL_C_1}
    \end{align}
    with Fermi momentum $k_F=\pi N/L$.
		
    To relate the results for the effective LL model to numerical results of the $J$-$V$ model at half filling, we use 
    Bethe ansatz results obtained via a mapping  to the spin-$1/2$ XXZ chain \cite{Giamarchi.2010}
    \begin{align}
        K &\equiv  \sqrt{\frac{v_F+g_4+g_2}{v_F+g_4-g_2}} 
        = \frac{\pi}{2\cos^{-1}\left(\frac{-V}{2J}\right)}\\
        \frac{v}{J} &= \frac{1} 
        {1-(2K)^{-1}} \sin\left[\pi(1-(2K)^{-1})\right] \ ,
    \end{align}
    where $K$ is the LL interaction parameter, and $v|q|$ is the dispersion relation for low energy excitations. We   
    use the above expressions  for $v$ and $K$ in the  diagonalized version of the  the LL Hamiltonian Eq.~\eqref{Eq:eq_LL_Hbos} to parametrize the interaction strength and velocity.

    The Hamiltonian Eq.~\eqref{Eq:eq_LL_Hbos} can be diagonalized using a Bogoliubov transformation 
    \begin{align}
      \begin{split}
        a_q  &= \cosh(\theta_q) b_q+\sinh(\theta_q)b_{-q}^\dagger \\
        a_{-q}^\dagger &= \sinh(\theta_q)b_q+\cosh(\theta_q)b_{-q}^\dagger \ ,
      \end{split} \label{Eq:eq_LL_bogTrafo}
    \end{align}
    in contrast  to simply  diagonalizing the Hamiltonian in a basis $(b^\dagger_q,b_{-q})$ (as one would do for a fermionic BCS Hamiltonian), since this would not preserve the bosonic commutation relations \cite{Bogo58,DelMaestro.2004}.  
    The choice of coefficients in Eq.~\eqref{Eq:eq_LL_bogTrafo} guarantees bosonic commutation relations $[a_q,a_{q'}^\dagger]=\delta_{q,q'}$, $[a_q,a_{q'}]=0$, $[a_q^\dagger,a_{q'}^\dagger]=0$, and one finds that 
    \begin{align}
        &\sum_q f(|q|)\frac{a_q^\dagger a_q}{\cosh^2(\theta_q)+\sinh^2(\theta_q)} \notag\\
        &= \sum_q f(|q|) b_q^\dagger b_q  + f(|q|)\frac{\sinh(\theta_q)\cosh(\theta_q)}{\sinh^2(\theta_q)+\cosh^2(\theta_q)}  
        \notag\\
        &\phantom{= \sum_q   \ }
        \times \left(b_qb_{-q}+b_q^\dagger b_{-q}^\dagger\right)\ .
    \end{align} 
    Choosing $f(|q|)=\omega_0(q)+m(q)$ and  $\tanh(2\theta_q)=g_2(q)/f(|q|)$, which in the limit $q\rightarrow0$ is given by $g_2/(v_F+g_4)$,  the Hamiltonian becomes diagonal
    \begin{align}
        H &= \sum_q \omega(q) a_q^\dagger a_q \\
        \omega(q) &=\sqrt{(\omega_0(q)+m(q))^2-g_2(q)^2}
        \equiv v |q|\ .
    \end{align}
    This allows us to evaluate the ground state expectation values
    \begin{align} 
        \ex{a_q^\dagger a_{q'}} &=    \delta_{qq'}f_b(q)  
        \label{Eq:eq_LL_exada}\\
        \ex{a_q a_{q'}} &= 0 = \ex{a_q^\dagger a_{q'}^\dagger} \ . \label{Eq:eq_LL_exaaadad}
    \end{align}
    where $f_b(q)$ is the Bose-Einstein distribution function with energies $\omega(q)$. 
 
    Using Eq.~\eqref{Eq:eq_LL_psi} in Eq.~\eqref{Eq:eq_LL_obdm_1}  together with the Baker-Hausdorff formula $e^A e^B = e^{A+B} e^{[A,B]/2}$, the one point correlation function becomes
    \begin{align}
        C_\alpha(x,0) &= \frac{1}{2\pi \eta}  e^{\alpha \frac{\pi x}{L} [N_\alpha, \varphi_{0,\alpha}]} e^{\frac{1}{2} [\phi_\alpha(x),\phi_\alpha(0)]} 
        \notag\\
        & \ \ \ \ \ \times \ex{e^{\imath (\phi_\alpha(x)-\phi_\alpha(0))}} \ . 
        \label{Eq:eq_LL_Calp}
    \end{align}
    Here, we use the boson cummulant formula $\ex{e^{\imath (\phi_\alpha(x)-\psi_\alpha(0))}} = e^{-\frac{1}{2}\ex{ (\phi_\alpha(x)-\psi_\alpha(0))^2}}$, 
    which is valid in equilibrium for a quadratic Hamiltonian, for any linear combination of bosons $\sum_n A_nb_n+B_nb^\dagger_n$. 
     In addition, 
    \begin{align}
        &\frac{1}{2} [\phi_\alpha(x),\phi_\alpha(0)] \notag\\
        &= 
        \frac{1}{2}\sum_{q>0}\frac{2\pi}{q L} \Big[
        e^{-q \imath (-\imath\eta-\alpha x)}  
        - e^{-  \imath q (-\imath\eta+\alpha x)}
        \Big] \ . \label{Eq:eq_LL_phicom}
    \end{align}
    Due to the regularization $\eta$, we can perform the $q$ sum, where we use that for any complex number $z$ with ${\rm Im}[z]>0$ holds \cite{Eggert.2009}
    \begin{align}
        \sum_{q>0} \frac{2\pi}{qL}e^{-\imath q  z} 
        &= -\ln\left[1-e^{-\imath\frac{2\pi}{L}z} \right]\\
        &= -\ln\left[2\imath e^{-\imath\frac{\pi}{L}  z} \sin\left(\frac{\pi}{L}z\right)\right] \ . \label{Eq:eq_LL_qsum}
    \end{align}
    One needs to be careful when using logarithm laws with complex numbers, as we need to stay on the main branch of the logarithm: $\ln(z)=\ln(|z|)+\imath\arg(z)$. With this in mind, we find for Eq.~\eqref{Eq:eq_LL_phicom}
    \begin{align}
        &\frac{1}{2}[\phi_\alpha(x),\phi_\alpha(0)] \notag\\
        &= \frac{1}{2}\Big\{
            -\ln\left[2\imath e^{-\imath\frac{\pi}{L} (-\alpha x -\imath\eta)}\cord{\alpha x + \imath \eta}\right] 
            \notag\\
        &\phantom{=\;\;}
            +\ln\left[-2\imath  e^{\imath\frac{\pi}{L} (-\alpha x +\imath\eta)}\cord{\alpha x - \imath \eta)}\right]
        \Big\}\\
        &=-\frac{1}{2}\ln\left|\frac{\cord{\alpha x + \imath\eta}}{\cord{\alpha x - \imath\eta}}\right| 
        \notag\\
        &\phantom{=\;}-\imath\arg\left[\cord{\alpha x+\imath\eta}\right]  
        -\imath\frac{\pi}{2} - \imath\frac{\pi}{L}\alpha x \ \label{Eq:phiphi_comm}.
    \end{align} 
    In the limit $\eta/x\rightarrow 0$, the $\arg$ term is 0 for $\alpha x>0$ and $\pm\pi$ if $\alpha x<0$ such that
    \begin{align}
         \lim_{\eta/x\rightarrow 0} e^{\frac{1}{2}[\phi_\alpha(x),\phi_\alpha(0)]}
         &= -\mathrm{sgn}(\alpha x) \imath e^{-\imath\frac{\pi}{L} \alpha x} \ .
         \label{Eq:eq_LL_ecomphi}
    \end{align}
    In order to evaluate the expectation value in Eq.~\eqref{Eq:eq_LL_Calp} by using the boson cummulant formula, we need the expectation values $\ex{\phi_\alpha(x)\phi_\alpha(x')}$. To compute them by utilizing the expectation values Eq.~\eqref{Eq:eq_LL_exada}, we  
    insert the  inverse of the transformation  Eq.~\eqref{Eq:eq_LL_bogTrafo} into the expression for $\phi_\alpha(x)$, Eq.~\eqref{Eq:eq_LL_phi}, such that  
    \begin{align}
        &\phi_\alpha(x)= -\sum_{q>0} \sqrt{\frac{2\pi}{qL}} e^{-q \eta/2}  \notag\\
        & \ \ \ 
        \times\Big[
            e^{\imath\alpha qx}\left(\cosh(\theta_q) a_q - \sinh(\theta_q) a_{-q}^\dagger\right)
             \notag\\
        &   \ \ \ \phantom{\times\Big[} 
            + e^{-\imath\alpha qx}\left(\cosh(\theta_q) a^\dagger_q - \sinh(\theta_q) a_{-q}\right)
        \Big] \ .
    \end{align}
    Using the expectation values of pairs for $a_q$ operators with $\ex{a_q^\dagger a_{q'}} = \delta_{qq'}f_b(q)$, we find 
    \begin{align}
        &\ex{\phi_\alpha(x)\phi_\alpha(x')}  
         = \sum_{q>0}\frac{2\pi}{qL} e^{-q\eta}
        \notag\\
        &\times \Big\{ e^{\imath\alpha (x-x')} \left[(1-f_b(q))\cosh^2(\theta_q) + f_b(q)\sinh^2(\theta_q)\right] 
        \notag\\
        &\phantom{  \Big[ }
        + e^{-\imath \alpha(x-x')} \left[f_b(q)\cosh^2(\theta_q) + (1-f_b(q))\sinh^2(\theta_q)\right]\Big\}.
              \end{align}
    At zero temperature, the Bose-Einstein distribution becomes $f_b(q>0) =0$, such that we find for the exponent appearing in the correlation function
    \begin{align}
        &-\frac{1}{2}\ex{(\phi_\alpha(x)-\phi_\alpha(0))^2}
        \notag\\
        &\phantom{--}
        = \frac{1}{2}\sum_{q>0}\frac{2\pi}{qL} e^{-q\eta} \left(\cosh^2(\theta_q)+\sinh^2(\theta_q) -1 +1\right)
        \notag\\
        &\phantom{---=\,\sum_{q>0}}
        \times\left[-2+e^{\imath\alpha qx} + e^{-\imath\alpha q x}\right] \ .
        \label{Eq:eq_LL_phiphiexp}
    \end{align} 
    Here, we added a zero ($-1+1$) to separate the free term Eq.~(\ref{Eq:eq_LL_qsum}) from the interaction term of the correlation function. Including the $-1$ in the interaction term ensures that it vanishes in the non-interacting case where $\theta_q\to 0$. We now precisely define the interaction cutoff $\eps$ by using it to describe the $q$ dependence of the interaction term as \cite{Iucci:2009lg}
    \begin{align}
        \cosh^2(\theta_q)+\sinh^2(\theta_q) -1 
        &\approx \frac{K+K^{-1}-2}{2} e^{-\eps |q|} \\
        &\equiv \gamma_{\rm eq}^2 e^{-\eps|q|} \ , \label{Eq:eq_LL_gammaeq}
    \end{align}
    where now $K=\lim_{q \to 0} e^{2\theta_q}$ and $\gamma_{\rm eq}$ are independent of $q$. While $\eps$ appears to be a free parameter of the model, we will show later that for not too strong interactions, a fixed  value can be chosen such that   the analytic calculation 
    reproduces  numerical results from exact diagonalization and DMRG for a range of interaction strengths. In addition, $\eps$ regularizes the interaction part of the correlation function and therefore allows us to take the limit  $\eta q \to 0 $ when keeping  $\varepsilon q $ finite.
     
    At zero temperature, we use Eq.~\eqref{Eq:eq_LL_qsum} to perform the $q$ sums in  Eq.~\eqref{Eq:eq_LL_phiphiexp}, and find
    \begin{align}
        &F_\alpha^0(x;\eta) \equiv
        -\frac{1}{2}\ex{(\phi_\alpha(x)-\phi_\alpha(0))^2}_{0}  
        \notag\\
        &\ = \frac{1}{2}\sum_{q>0}\frac{2\pi}{qL}e^{-\imath q(-\imath\eta)} 
        \left[-2+e^{\imath\alpha qx}+e^{-\imath\alpha qx}\right] \\
        &\ = \ln\left[-2\imath e^{-\frac{\pi}{L}\eta}     \sin\left(\frac{\pi}{L}\imath \eta\right)\right]
        \notag\\&\phantom{=\,}
             - \frac{1}{2}\ln\left[2\imath e^{-\imath\frac{\pi}{L}(\alpha x-\imath\eta)}
            \cord{\alpha x-\imath\eta}\right]
        \notag\\&\phantom{=\,} 
            -\frac{1}{2}\ln\left[2\imath e^{-\imath\frac{\pi}{L}(\alpha x+\imath\eta)}
            \cord{\alpha x+\imath\eta}\right] \ ,
            \label{Eq:eq_phiphi_free}
    \end{align}
    such that the interaction term can be obtained from the free one by multiplying with a   factor $\gamma_{\rm eq}^2$ while  changing the regularization to include the interaction cutoff, i.e. $\eta \to \eta+\eps$,
    \begin{align}
        &-\frac{1}{2}\left[\ex{(\phi_\alpha(x)-\phi_\alpha(0))^2}-\ex{(\phi_\alpha(x)-\phi_\alpha(0))^2}_0\right]
        \notag\\ 
        & \phantom{=\,}
        = \gamma_{\rm eq}^2 F_\alpha^0(x;\eta+\eps)    \ .
    \end{align} 
    We use this and the translational invariance of the expectation value, i.e. $\ex{\phi_\alpha(x)\phi(x')}=\ex{\phi_\alpha(x-x')\phi(0)}$, to obtain the expectation value that appears in the one point correlation function
    \begin{align}
        &e^{-\frac{1}{2}\ex{(\phi_\alpha(x)-\phi_\alpha(0))^2}}
        \notag\\ 
        &= \frac{-\imath\sin\left(\frac{\pi}{L} \imath \eta\right)}{\left|\cord{\alpha x + \imath\eta}\right|} \left[ \frac{\imath \sin\left(\frac{\pi}{L} \imath (\eta+\eps)\right)}{  |\cord{\alpha x + \imath(\eta+\eps)}| } \right]^{\gamma_{\rm eq}^2} .\label{Eq:eq_LL_bcf}
    \end{align}
    Using Eq.~\eqref{Eq:eq_LL_ecomphi} and Eq.~\eqref{Eq:eq_LL_bcf} in the expression for the correlation function Eq.~\eqref{Eq:eq_LL_Calp}, taking the limit $\eta/x,\eta/L\rightarrow0$, where $\sin(\pi \imath\eta/L)/\eta\rightarrow \pi \imath/L$,  $\imath\sin(\imath b)=-|\sin(\imath b)|$, and $\sqrt{\sin(b+\imath c)\sin(-b+ \imath c)}=\imath |\sin(b+\imath c)|$, we find 
    \begin{align}
        C_\alpha(x,0)&= \frac{\imath\pi}{2\pi L} \frac{\mathrm{sgn}(\alpha x)}{\left|\sin\left(\frac{\pi}{L}(\alpha x)\right)\right|} \left|\frac{\sin\left(\frac{\pi}{L} \imath\eps\right)}{\cord{\alpha x +\imath \eps}}\right|^{\gamma_{\rm eq}^2}\\
        &= \frac{\alpha \imath}{2\sin(\pi x/L)} \left| \frac{\sin(\pi \imath\eps/L)}{\cord{x+\imath\eps}} \right|^{\gamma_{\rm eq}^2}.
    \end{align}
     
\begin{figure}[ t! ]
			\centering
			\includegraphics[width=8.6cm]{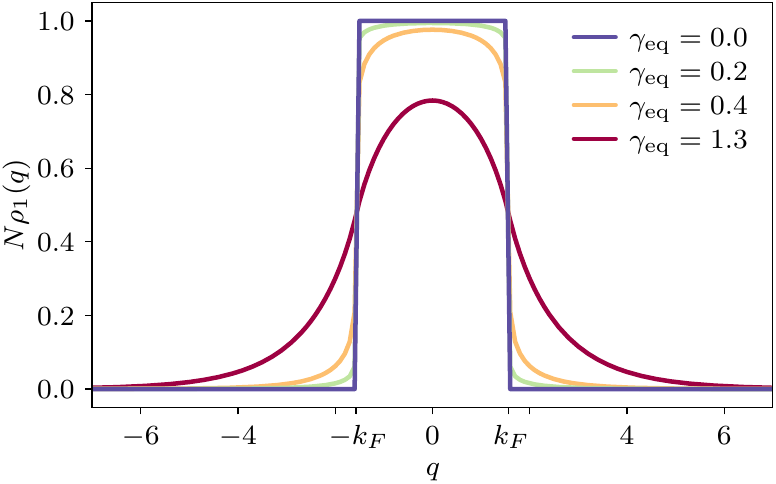}  
\caption{\label{Fig:eq_LL_fq}Distribution function $N\rho_1(q)$, Eq.~\eqref{Eq:eq_LL_fq}, obtained from the $x/L\ll 1$ limit of the one body reduced density matrix in the Luttinger liquid model for a fixed interaction cutoff $\eps=0.84$ and various interaction strengths $V/J$. Without interactions, $\gamma_{\rm eq}=0$, the distribution function is a step function up at the Fermi momenta $k_F=\pm \pi/2$.}
\end{figure}

    Using that $\alpha=-1$ for left movers and $\alpha=+1$ for right movers, we obtain the full one body density matrix Eq.~\eqref{Eq:eq_LL_obdm_1} 
    \begin{align} 
        \rho_1(x,0) &=  \rho_1^0(x,0)\,\left|
                   \frac{ \sin(\pi \imath \eps/L)}{ \cord{x+\imath  \eps}}
                  \right|^{\gamma_{\rm eq}^2}  
                  \label{Eq:eq_LL_rhoTDlimit}\\
        \rho_{1}^0(x,0) &= \frac{1}{N}\frac{\sin(k_F x)}{L\sin(\pi x/L)}  \ .
        \label{Eq:eq_LL_rho0TDlimit}
    \end{align} 
    We show in Appendix \ref{App:latticeGF} that Eq.~\eqref{Eq:eq_LL_rho0TDlimit} is equivalent to the exact  one body density matrix for non-interacting lattice fermions.
    Because $x$ is a relative coordinate and we are interested in the short distance behavior that dominates the Fourier transform and the entropy, we consider the limit of large $L$ with $x/L\ll 1$ and neglect terms of order $\mathcal{O}(x/L)$. For the leading term $L\sin(\pi x/L)\rightarrow \pi x$  we then arrive at the following expression for the one body density matrix:
    \begin{align}
        \rho_{1}(x,0) &=  \frac{\sin(k_F x)}{N \pi x} \left( \frac{\eps^2}{x^2 + \eps^2} \right)^{\gamma_{\rm eq}^2/2} 
        +\mathcal{O}\left(\frac{x}{L}\right)\ , 
        \label{Eq:eq_LL_fqTDlimit}
    \end{align}
    which is normalized such that $L\rho_1(0,0)=1$ where $k_F=\pi N/L$. Because the particle number $N$ appears explicitly in the normalization, we cannot directly take the thermodynamic limit. Therefore, we first compute the entropy density,  and only then take 
    the thermodynamic limit $1/N\rightarrow 0$.  To diagonalize $\rho_1(x,0)$, we compute the Fourier transform which yields  
    \begin{align}
        \rho_1(q) &= \int_{-\infty}^\infty dx\, \rho_1(x,0)e^{-\imath qx} \notag\\
         &= \ \frac{\Gamma[\frac{1}{2} (-1+\gamma_{\rm eq}^2)] \sqrt{\pi}}{ 2\pi N \Gamma(\gamma_{\rm eq}^2/2)}  \notag
      \left[ f_1(\tilde{q}) + f_1(-\tilde{q}) \right] \\
            &\ \ - \frac{2\Gamma(-\gamma_{\rm eq}^2) \sin(\pi\gamma_{\rm eq}^2/2)}{2\pi N} \, \left[ f_2(\tilde{q}) + f_2(-\tilde{q})\right]  \label{Eq:eq_LL_fq}
    \end{align}
    where $\tilde{q} = \eps q$, $\tilde{k}_F = \eps k_F$, $L/(2\pi)\int dq\,\rho_1(q)=1$, and  
    \begin{align*}
        f_1(\bar{q})&= (\tilde{k}_F+\tilde{q}) \, _{1}F_{2}\left[\left\{ \frac{1}{2}\right\},\left\{ \frac{3}{2},\frac{3-\gamma_{\rm eq}^2}{2}\right\}, \frac{1}{4} (\tilde{k}_F + \tilde{q})^2\right]\\
        f_2(\tilde{q})&= (\tilde{k}_F+\tilde{q}) |\tilde{k}_F+\tilde{q}|^{\gamma_{\rm eq}^2- 1} 
        \notag\\ 
        &\times\,_{1}F_{2}\left[\left\{ \frac{\gamma_{\rm eq}^2}{2}\right\},\left\{ \frac{1+\gamma_{\rm eq}^2}{2},\frac{2+\gamma_{\rm eq}^2}{2}\right\}, \frac{1}{4} (\tilde{k}_F + \tilde{q})^2\right]  .
    \end{align*} 
    Here, ${_p}F_q$ are the generalized hypergeometric functions. From the Fourier transformed  one body density matrix $\rho_1(q)$ 
    we obtain the fermionic distribution function as  $N\rho_1(q)$, which in the absence of interactions $\gamma_{\rm eq}=0$  reduces to a step function $\theta(|q|-k_F)$, and in presence of interactions decays like a power law (see Fig.~\ref{Fig:eq_LL_fq}).
    
    % Entropy definitions
    Using that the 1-RDM is diagonal in Fourier space, we can directly compute the one-particle \ren entanglement entropy  
    \begin{align}
       S_\alpha&= \frac{1}{1-\alpha}\ \ln\left(\frac{N}{2k_F}\int dq \; \rho_1(q)^\alpha \right) \label{Eq:eq_LL_Renyis}\\
       S_1 &=  -\frac{N}{2k_F}\int dq\; \rho_1(q)\ln \rho_1(q)   \ , \label{Eq:eq_LL_vonNeumann}
    \end{align}
    where $\alpha=1$ is the von Neumann entropy, and the factor $N/(2k_F)=(2\pi/L)^{-1}$ originates from turning the sums into integrals in the limit of large $L$. When comparing to numerical results, we additionally subtract the entropy for free fermions $S_{\rm ff}$. 

In the absence of interactions, $\gamma_{\rm eq}=0$, the one-body density is given by $\varrho_1^{0}(x,0)$ in Eq.~\eqref{Eq:eq_LL_rho0TDlimit}. Performing the Fourier transform, we recover the expected zero temperature distribution function 
    \begin{align}
        N\varrho_1^0(q) = \theta(|q|-k_F) \ .
    \end{align}
    Using this expression in  Eq.~\eqref{Eq:eq_LL_vonNeumann}, one finds the
    free fermion von Neumann entropy
    \cite{yang1962concept,Carlson:1961dr,coleman1963structure,sasaki1965eigenvalues,lemm2017entropy,Lemm:2017ha,Cheong:2004la} 
    \begin{align}
        S_{\rm ff} &= -\frac{N}{2k_F}\int_{-k_F}^{k_F} dq \, \frac{1}{N}\ln\frac{1}{N}
         = \ln(N)\ .
    \end{align}
    The same one-particle entanglement entropy is obtained for any other \ren power $\alpha>1$ \cite{lemm2017entropy}, which can be seen from Eq.~\eqref{Eq:eq_LL_Renyis}
    \begin{align}
        S_{\alpha,\rm ff} &= \frac{1}{1-\alpha} \ln \left(\frac{N}{2k_F}\int_{-k_F}^{k_F} dq \, \frac{1}{N^\alpha}\right) \notag\\
        &= \frac{1}{1-\alpha}\ln\frac{1}{N^{\alpha-1}}      \notag  \\
        &= \ln(N) =S_{\rm ff}\ .
    \end{align}

% -------------------------------------------------------------
% -------------------------------------------------------------
   \section{Numerical Results for Equilibrium 1-Particle Entanglement}
   \label{sec:eqm_results}
    % Numerics
    To study how well the low energy field theory approach describes the $J$-$V$ model in Eq.~\eqref{Eq:eq_JVHam} in the LL phase and to fix the interaction cutoff, we perform a series of numerical calculations on finite sized systems. The software needed to reproduce all results is open source and has been made available online \cite{CODE}.
   
    % ED equilibrium
    We first utilize exact diagonalization (ED), where we construct all $2N\choose N$ basis states for a lattice with $\Lb=2N$ sites and $N$ fermions to determine the corresponding matrix elements of Eq.~\eqref{Eq:eq_JVHam} and construct the Hamiltonian as a sparse matrix. We then use the Lanczos algorithm \cite{Lanczos:1950yo} to determine the ground state $\ket{\Psi_{0}}$, from which the full density matrix can be determined as ${\rho} = \ket{\Psi_0}\bra{\Psi_0}$. The reduced one-body density matrix is obtained by fixing one coordinate in the anti-symmetrized many particle wave function $\Psi_0(i_1,...,i_N)=\braket{i_1,...,i_N}{\Psi_0}$ and tracing out the other $N-1$ particle positions \cite{DelMaestro:2021ja}
    \begin{align}
       {\rho}_1^{i_1,j_1} = \sum_{\substack{i_2,...,i_N\\j_2,...,j_N}}\Psi^*_0(i_1,...,i_N)\Psi_0(j_1,...,j_N) \ .
    \end{align}

    As a second numerical approach, we consider approximate methods that allow us to consider much larger systems. For this, we obtain the ground state $\ket{\Psi_0}$ using DMRG, and the implementation of states as matrix product states (MPS) in \texttt{ITensors.jl} \cite{itensor} allows to directly compute the reduced one body density via
    \begin{align}
    	{\rho}_1^{i_1,j_2} &= \frac{1}{N} \bra{\Psi_0} c_{i_1}^\dagger c_{j_1} \ket{\Psi_0}\ ,
    \end{align}
where $c_i^\dagger$ is the creation operator on lattice site $i$.
    
From the reduced density matrix, we compute the one-particle \ren entanglement entropy for \ren index $\alpha$ using 
\begin{equation}
       S_\alpha = \frac{1}{1-\alpha}\, \ln(\Tr[{\rho}_1^\alpha])\, , \label{Eq:eq_LLRenyi}
\end{equation}
where the von Neumann entropy is obtained as the limit $\alpha \to 1$:
\begin{equation}
       S_1 = \Tr[{\rho}_1\ln({\rho}_1)] \ .
\end{equation}

\subsection{Symmetry Decomposition of the Lattice Hamiltonian}
\label{subsec:symmetries}

    While ED provides approximation-free access to the ground state, scaling of the size of the Hamiltonian $\propto {2N\choose N}$ makes it prohibitive to consider systems with $L\gtrsim 40$. Using a series of optimizations, we are able to compute one particle entanglement entropies with ED for systems with up to $N=19$ fermions with 1TB of system memory. The crucial factor for reducing the complexity of the problem is the use the symmetries of the Hamiltonian Eq.~\eqref{Eq:eq_JVHam}, which we define below by their action on the occupation numbers of the states. We discuss the action of the symmetry operators on the fermion operators $c_i^{\phantom{\dagger}}, c_i^\dagger$ in greater detail in Appendix \ref{App:SymOperators}. 
   
\subsubsection{Translation symmetry}
\label{ssec:translational_symmetry}

Translation symmetry $T$  
which moves each fermion one site to the right, \emph{e.g.} $T\ket{011001}=\ket{101100}$,
    commutes with the Hamiltonian, $[H,T]=0$, due to the boundary conditions, which allows us to group basis states in symmetry cycles such that each state of a cycle $\nu$ is mapped onto another state in the same cycle by $T$. Choosing one state $\ket{\varphi_\nu}$ from each cycle, the so-called cycle leader, we can define new basis states as linear combinations
    \begin{align}
       \ket{\phi_{\nu,q}} &=\frac{1}{\sqrt{M_\nu}} \sum_{m=1}^{M_\nu} 
       e^{\imath\frac{2\pi q}{M_\nu}\,(m-1)} T^{m-1} \ket{\varphi_\nu}\ ,
       \label{Eq:eq_ED_symCycles}
    \end{align}
where $M_\nu$ is the length of cycle $\nu$ with $T^{M_\nu}=1$ within the cycle.  Because $T$ commutes with $H$, the Hamiltonian becomes block diagonal when sorting the basis states according to the values of $q$. The main advantage of using this basis is that the ground state always lies in the $q=0$ block \cite{Barghathi:2022rg}, and it is therefore sufficient to compute and store only this single block, reducing the size of the required basis roughly by a factor of $1/L$.
  
\subsubsection{Particle hole symmetry}
\label{ssec:particle_hole_symmetry}

At half filling, the particle-hole operator $P$, which flips all occupation numbers $P\ket{101001}=\ket{010110}$
is another symmetry of the Hamiltonian which also commutes with the translation operator, $[T,P]=0$. Because $P^2=1$, the particle-hole operator has eigenvalues $n_P=\pm1$. If $P\ket{\phi_{\tilde{\nu},q}}$ lies in a different cycle  than $\ket{\phi_{\tilde{\nu},q}}$, we can use $P$ to further subdivide the $q=0$ block by using the projection $(1\pm P)/\sqrt{2}$ onto its eigenstates
    \begin{align}
       \ket{\theta_{\tilde{\nu},q,n_P=\pm1}} = \frac{1}{\sqrt{2}} 
          \left( \ket{\phi_{\tilde{\nu},q}} \pm P\ket{\phi_{\tilde{\nu},q}}  \right)\ . \label{Eq:eq_ED_symPstates}
    \end{align}

\subsubsection{Reflection symmetry}
\label{ssec:reflection_symmetry}

The third symmetry we exploit is spatial inversion $R$, which reflects the occupation numbers $R\ket{011011}=\ket{110110}$ about a site and commutes with the Hamiltonian $[R,H]=0$. However, in general, $R$ does not commute with $T$, but fortunately in the $q=0$ block translation and spatial inversion do commute. Since $R^2=1$, the eigenvalues are also given by $n_R=\pm 1$ and the projection operator is $(1\pm R)/\sqrt{2}$. If $R$ maps either $\ket{\phi_{\nu,q}}$ or $\ket{\theta_{\tilde{\nu},q,n_P}}$ into another cycle, projecting onto eigenstates of $R$ further subdivides the $q=0$ block of the Hamiltonian in analogy to Eq.~\eqref{Eq:eq_ED_symPstates}. 
   
    %Single block needed
We therefore only need to construct the $q=0$, $n_R=+1$, $n_P=+1$ block of the Hamiltonian which is a major reduction in memory and time complexity for obtaining the ground state. We can further use translation symmetry in a similar way to reduce the computational effort when computing the reduced density matrix from the ground state. In addition, for our ED implementation, we encode states using a 64bit integer basis, where each bit of the binary representation of an integer represents the occupation number of the site at the corresponding position \cite{Lin:1990gu}.  This has the advantage that symmetry operation can be implemented very efficiently using low-level bit operations and that we avoid all overhead of using vectors containing the occupation numbers for each state.
    
    % DMRG
\subsection{Density Matrix Renormalization Group}
\label{subsec:dmrg}
In order to study systems with $L \ge 40$ and thus improve finite size scaling to the thermodynamic limit, we additionally use the DMRG implementation of the \texttt{ITensors.jl} software package \cite{itensor} for the Julia programming language. As an approximate method, DMRG does not need to explore the entire Hilbert space and therefore requires fewer resources, but at the price of inaccuracies with magnitudes that are difficult to estimate a priori.  We therefore also use ED as a benchmark to estimate the reliability of DMRG results, where a direct naive DMRG application to the $J$-$V$ model with periodic boundary conditions leads to significant errors already for systems of size $N>17$. We thus use a number of checks and detailed knowledge of the physical system to stabilize the DMRG calculation. 
    
\subsubsection{Initial state}
\label{ssec:initial_state}

    It is crucial to construct very good initial states, so that the algorithm starts as close as possible to the ground state. For this purpose, we combine a state which is a superposition of random states of the correct particle number, with a $V/J$ dependent fraction of the corresponding $\ket{\Psi_{V/J\rightarrow\pm \infty}}$ state (Eq.~\eqref{Eq:eq_JV_psiPinf}, Eq.~\eqref{Eq:eq_JV_psiMinf}).
    
\subsubsection{Orthogonal subspace}
\label{ssec:orthogonal_subspace}
The most important step for stabilizing convergence of DMRG to the ground state is to construct an orthogonal subspace to the ground state and enforce orthogonality to a basis in this subspace during each sweep of DMRG. This feature has already been implemented in \texttt{ITensors.jl} with the intent to obtain excited states. The consideration of symmetry cycles Eq.~\eqref{Eq:eq_ED_symCycles} already reveals good candidates for orthogonal subspaces, because states with different $q$ are orthogonal to each other. Using all states from blocks $q>0$ is overkill, slowing down the DMRG algorithm and requiring huge amounts of memory, eliminating the advantages of the approximation method. We therefore only consider a subspace in which DMRG is most likely to converge if it misses the ground state. For $V/J>0$ using the two states with maximal particle separation $\ket{\psi_{>,\nu}}=T^\nu(\prod_{j=1}^Nc_{2j}^\dagger)\ket{0}$ this is:
    \begin{align}
        \ket{\Psi_{\bot,>}} &= \frac{1}{\sqrt{2}} \left[\ket{\psi_{>,0}}-\ket{\psi_{>,1}} \right]  \ ,
        \intertext{and for negative interactions there are $L$ states with full clustered fermions $\ket{\psi_{<,\nu}}=T^\nu(\prod_{j=1}^Nc_j^\dagger)\ket{0}$. Their span is given by }
        \ket{\Psi_{\bot,<,q}}  &= \frac{1}{\sqrt{N}} \sum_{\nu=0}^{L-1}
            \cos\left(\frac{2\pi\nu q}{L}\right)\,\ket{\psi_{<,\nu}}\\
        \ket{\Psi_{\bot,<,q+N}}  &= \frac{1}{\sqrt{N}} \sum_{\nu=0}^{L-1}
            \sin\left(\frac{2\pi\nu q}{L}\right)\,\ket{\psi_{<,\nu}} \ .
    \end{align}
    
    \begin{figure}[ t! ]
			\centering
			\includegraphics[width=8.6cm]{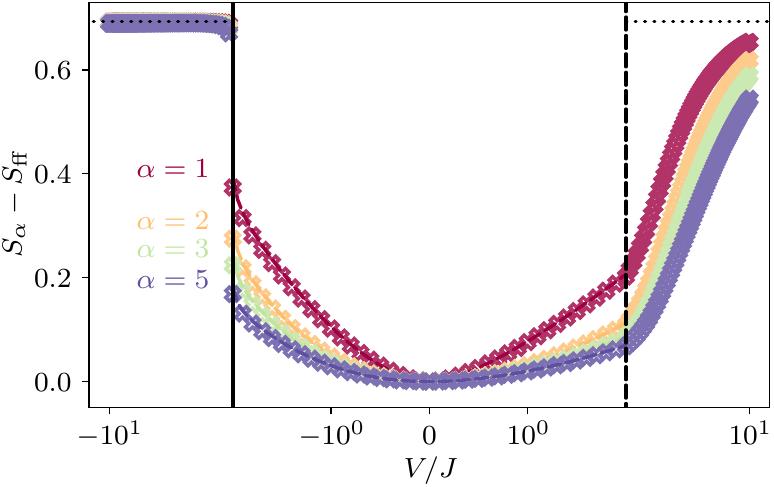}  
			\caption{\label{Fig:eq_DMRG_N51}One-particle \ren entanglement entropy $S_\alpha$ for different values of the \ren index $\alpha$ as a function of the interaction strength $V/J$ where $S_{\rm ff}$ is the 1-particle entropy for free fermions. The crosses are obtained using DMRG for $N=51$  on a lattice of $\Lb=102$ sites. Solid lines depict extrapolation to the thermodynamic limit from ED and DMRG data, and dashed horizontal lines show theory predictions for $|V/J|\rightarrow\infty$. Phase transitions in the $J$-$V$ model are marked with vertical lines at $V/J=\pm 2$. }
	\end{figure}

\subsection{Ground State DMRG and ED Results}
\label{subsec:numerical_results}

% Describe DMRG N=51 figure
By forcing the ground state to be orthogonal to these states, we are able
to consider systems with sizes $N=51$ or larger at half filling with periodic boundary conditions. Fig.~\ref{Fig:eq_DMRG_N51}
shows the 1-particle entanglement entropy (crosses) calculated with DMRG
for $N=51$ for a large range of interaction strengths spanning all phases in the $J$-$V$ model.  For $V/J=-2$, the first order phase transition is clearly visible and
$S_\alpha-S_{\rm ff}$ remains stable, reaching the theoretical value
$\ln(2)$ \cite{Haque:2009zi} for large negative $V/J$. For free fermions, where $V/J=0$ in the LL phase, the one-particle entanglement entropy vanishes as expected \cite{Haque:2009zi}. Additionally, at the second-order phase transition into the charge density wave phase near $V/J=2$, a change in the slope of the entropy is visible, which then slowly approaches the theoretical value $\ln(2)$ \cite{Haque:2009zi}.

\begin{figure}[ t! ]
			\centering
			\includegraphics[width=8.6cm]{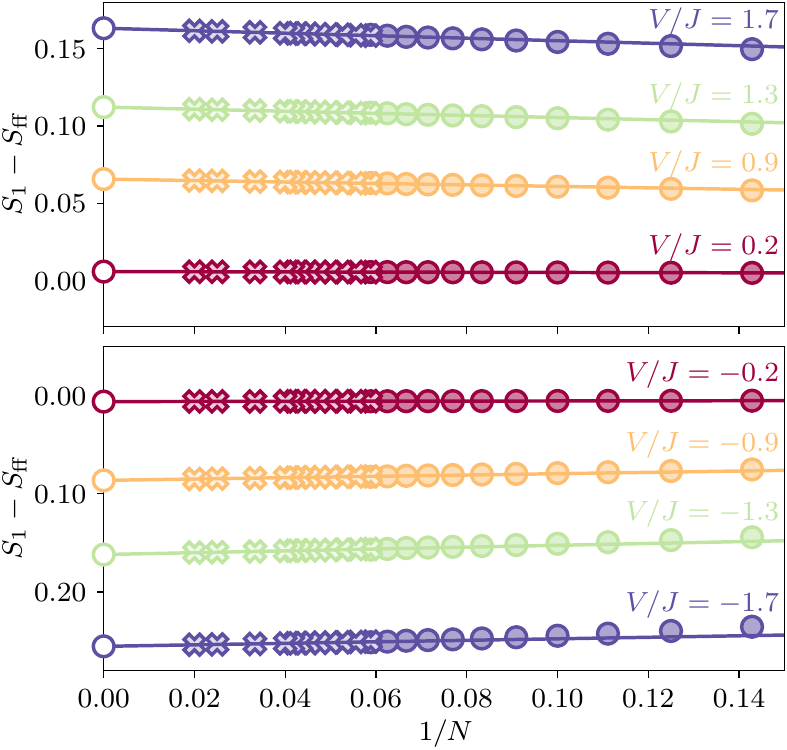}  
                        \caption{\label{Fig:eq_fsscaling}Finite size scaling of the equilibrium one-particle von Neumann entanglement entropy $S_1$ for various interaction strengths $V/J$ where the free fermion contribution $S_{\rm ff}$ has been subtracted. Results obtained with DMRG (crosses) and ED (circles) are shown together along with a linear extrapolation to the thermodynamic limit $1/N\rightarrow 0$. ED provides access to lattices at half filling with up to $N=19$ fermions and using DMRG lattices with more than $N=51$ fermions can be studied.}
	\end{figure}
	 
        For comparison with field theory, we first estimate the thermodynamic limit $N\rightarrow\infty$ by finite size scaling of the numerical results for the one-particle entanglement entropy, with a general scaling form introduced by Haque \emph{et al.} \cite{Haque:2009zi} and confirmed in subsequent works \cite{Herdman:2015xa,Barghathi:2017ab}:
	\begin{align}
	    S_\alpha(N,V/J) &= \ln(N) + A_\alpha(V/J)+\mathcal{O}(N^{-\lambda}) \ ,
	    \label{eq:Adef}
	\end{align}
with $\lambda>0$.  In subsequent figures, we will focus on the behavior of the constant correction $A_\alpha(V/J)$ to the leading order logarithmic scaling. 

For reliable finite size scaling, we calculate $S_\alpha-S_{\rm ff}$ for systems with $N=2,3,\dots,19$ fermions using ED and fermion numbers between $N=17$ and $N=51$ using DMRG and then extrapolate linearly to $1/N\rightarrow 0$ (white filled circles in Fig.~\ref{Fig:eq_fsscaling}). We find very good $1/N$ scaling and excellent agreement between exact ED (colored circles in Fig.~\ref{Fig:eq_fsscaling}) and approximate DMRG (crosses in Fig.~\ref{Fig:eq_fsscaling}) everywhere in the LL phase.
	\begin{figure}[ t! ]
			\centering
			\includegraphics[width=8.6cm]{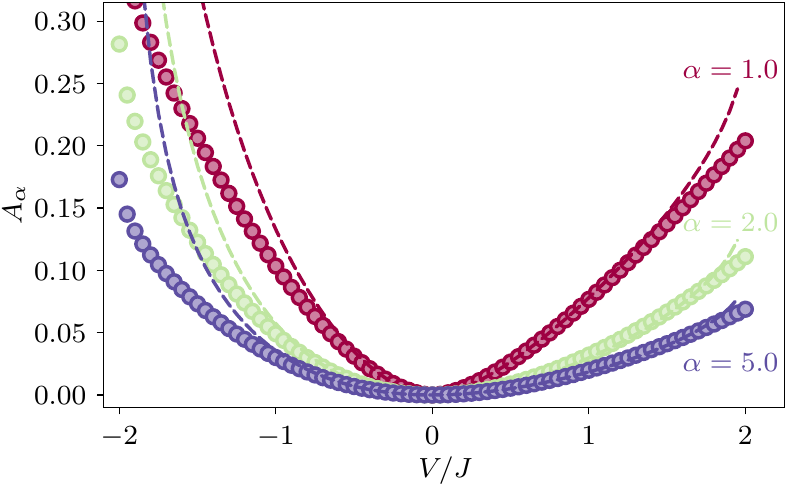}  
                        \caption{\label{Fig:eq_tdlimit_LL}Interaction strength $V/J$ dependence of the constant contribution to the 1-particle \ren entanglement entropies extrapolated to the thermodynamic limit $A_\alpha$ for \ren indices $\alpha=1, 2,$ and $5$ together with the prediction from bosonization for a fixed interaction cutoff $\eps=0.84$. We find very good agreement between the Luttinger liquid prediction and the numerical results for the $J$-$V$ model in region $V/J\in[-0.5,1.5]$.}
	\end{figure}

\subsection{Comparison with Luttinger Liquid Theory}
\label{subsec:comparison_LL_theory}

We perform this finite size scaling for all calculated interaction strengths $V/J$ and plot the constant contribution to the 1-particle entanglement entropies ($A_\alpha$, circles) as a function of $V/J$ in Fig.~\ref{Fig:eq_tdlimit_LL} along with the numerically integrated Luttinger liquid result from Eq.~\eqref{Eq:eq_LL_fq} and Eq.~\eqref{Eq:eq_LLRenyi} (dashed lines) for a fixed interaction cutoff $\eps=0.84$ determined via fitting.  We find excellent agreement between LL theory with this fixed cutoff and numerical results for the $J$-$V$ model for small to moderate interaction strengths $-0.5<V/J<1.5$. Close to the phase transitions and especially for large negative interaction strengths $V/J\rightarrow-2$, where $\gamma_{\rm eq}\rightarrow\infty$, significant deviations from the low energy LL theory are apparent. 

\begin{figure}[ t! ]
			\centering
			\includegraphics[width=8.6cm]{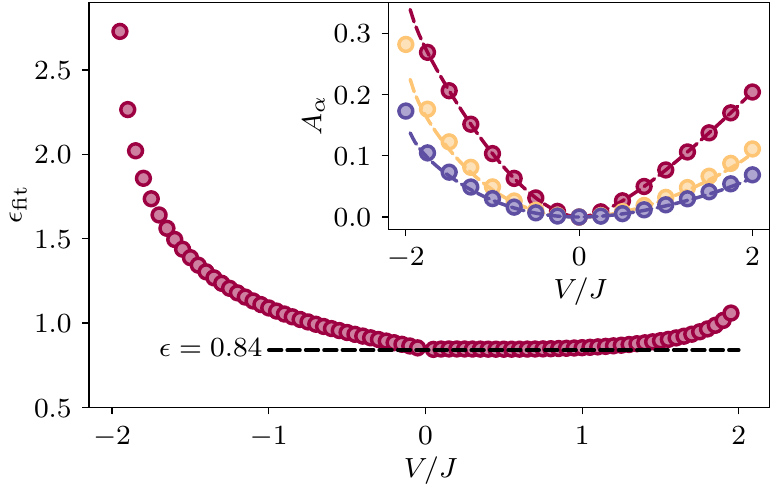}  
			\caption{\label{Fig:eq_fiteps}Effective interaction cutoff $\eps_{\rm fit}$ as a function of the interaction strength $V/J$ (red circles) obtained by fitting the Luttinger liquid prediction for the one-particle von Neumann entanglement entropy for each interaction strength individually to the numerical data of the $J$-$V$ model. We find an extended flat region of the effective cutoff $\eps=0.84$ (dashed, black line) that is nearly independent of interaction strength.
			The inset depicts numerical results for the \ren entropies with $\alpha=1$ (red circles), $\alpha=2$ (yellow circles), and $\alpha=5$ (blue circles) together with the field theory prediction using the fitted interaction dependent cutoff $\eps_{\rm fit}$.}
	\end{figure}

	% Fitting effective cutoff
To systematically study for which interactions the $J$-$V$ model can be accurately described by the LL model, we fit the bosonization prediction for each interaction strength individually to the finite size scaled data for the von Neumann entropy $A_1(V/J)$ as defined in Eq.~\eqref{eq:Adef}, to determine an effective interaction cutoff $\eps_{\rm fit}(V/J)$ (red circles in the main panel of Fig.~\ref{Fig:eq_fiteps}). We find an extended region with $\eps=0.84$ (dashed, black line) for small negative and positive interactions $V/J$ where the cutoff has minimal dependence on the interaction strength.  
With the obtained effective cutoff, we can fit the LL model at every point in the LL phase to the $J$-$V$ model with excellent agreement as shown in the inset of Fig.~\ref{Fig:eq_fiteps} where we plot the 1-particle entanglement entropies from numerics and for the effective interaction cutoff $\eps_{\rm fit}$. An interaction dependent cutoff for large interactions is also a consequence of approximating the $q$ dependence of the exponent $\gamma_{\rm eq}$ by the cutoff $e^{-\eps|q|}$ in field theory calculations (see Eq.~\eqref{Eq:eq_LL_gammaeq}) in order to make the $q$ sums analytically tractable.
 
\section{1-Particle entanglement entropy after a quantum quench}  

%J-V Hamiltonian for quench
    We next consider free fermions for $t<0$ and suddenly turn on the $V/J$ interaction at $t=0$, so that the $J$-$V$ Hamiltonian for this quench is given by 
    \begin{align}
        H &= -J \sum_{i=1}^{\Lb} (c_{i+1}^\dagger c_i^{\phantom{\dagger}} + c_i^\dagger c_{i+1}^{\phantom{\dagger}}) + V(t) \sum_{i=1}^{\Lb} n_i n_{i+1} \ , \label{Eq:qu_JVHam}
    \end{align}
    with $V(t)=\theta(t) V$. This allows us to study the growth and spread entanglement entropy after the quench by considering the difference $S_\alpha-\ln(N)$, in which the entropy of free fermions is subtracted.     
    % LL Hamiltonian for quench
    We again start by computing the one body density 
    \begin{align}
     \rho_1(x,0;t)&= \frac{1}{N} \qty[ e^{-\imath k_F x} C_+(x,0;t) + e^{\imath k_F x}C_-(x,0;t)] \\
        C_\alpha(x,0;t) &= \ex{\psi_\alpha^\dagger(x,t)\psi_\alpha(0,t)}
    \end{align}
    for the quench in the LL model 
    \begin{align}
        H &= \sum_{q} \left[\omega_0(q)+m(q,t)\right] b_q^\dagger(t) b_q(t) \notag \\
        & \ \ + \frac{1}{2} \sum_{q} g_2(q,t) \left(b_q(t) b_{-q}(t)+b_q^\dagger(t) b_{-q}^\dagger(t)\right) \label{Eq:qu_LL_Hbos} \ ,
    \end{align}  
    where in this case $g_2(q,t)=\theta(t) g_2(q)=\theta(t)g_2|q|$,  $m(q,t)=\theta(t)g_4|q|$, and again,  $\omega_0(q)=v_F|q|$. Analogous to the equilibrium case, we can diagonalize the Hamiltonian for any fixed time $t>0$ using the same Bogoliubov transformation Eq.~\eqref{Eq:eq_LL_bogTrafo} with   $\tanh(2\theta_q)=g_2(q)/(\omega_0(q)+g_4(q))$, but now the operators $a_q(t)$ are time dependent. For $t>0$ this yields the diagonal Hamiltonian
    \begin{align}
        H &= \sum_q v|q| a_q^\dagger(t)\,a_q(t)\\
        v &= \sqrt{(v_F+g_4)^2-g_2^2} \ . \label{Eq:qu_LL_vel}
    \end{align}
    Since the Hamiltonian for $t>0$ is diagonal in the $a_q$ operators, we can use the trivial time evolution
    \begin{align}
        a_q(t) &= e^{-\imath v |q| t} a_q \ .
    \end{align}
    Substituting this time evolution into the inverse of the transformation  Eq.~\eqref{Eq:eq_LL_bogTrafo}, we obtain the time evolution of the $b_q$ operators as \cite{Iucci:2009lg}
    \begin{align}
        b_q(t) &= w_q(t)\,b_q+u_q(t)\,b_{-q}^\dagger  \label{Eq:qu_bqTimeDep}\\ 
        \begin{split}
            w_q(t) &= \cos(v|q| t)-\imath\sin(v |q| t)\cosh(2\theta_q)\\
            u_q(t) &= -\imath\sin(v|q| t)\sinh(2\theta_q) \ .
        \end{split}
    \end{align}
    A very important conceptual difference to the equilibrium case is that the Hamiltonian is not diagonal in the $a_q$ operators for $t\rightarrow 0^-$, and therefore we cannot easily write down expectation values of the $a_q$ operators. However, since $H$ is diagonal in the $b_q$ operators for $t\rightarrow 0^-$,   we can use $b_q(t=0)\equiv b_q$ and  
    \begin{align}
      \begin{split}
        \ex{b_q^\dagger b_{q'}} &= f_b(q) \delta_{q,q'}  
        \\
        \ex{b_q b_{q'}} &= 0 =\ex{b_q^\dagger b^\dagger_{q'}} \ .
      \end{split}
    \end{align}
    This together with the more complicated time evolution of the $b_q$ operators Eq.~\eqref{Eq:qu_bqTimeDep} gives rise to a  different exponent $\gamma\geq\gamma_{\rm eq}$ as we show in the following.
    Up to Eq.~\eqref{Eq:eq_LL_ecomphi} the calculation for the correlation function is analogous to the equilibrium case such that  
    \begin{align}
        C_\alpha(x,t) &= \frac{e^{\alpha\frac{\imath\pi x}{L}}}{2\pi \eta}e^{\frac{1}{2}[\phi_\alpha(x,t),\phi_\alpha(0,t)]}e^{-\frac{1}{2}\langle(\phi_\alpha(x,t)-\phi_\alpha(0,t))^2\rangle} \ . \label{Eq:qu_LL_Calp}
    \end{align}
    The exponential $e^{\frac{1}{2}[\phi_\alpha(x,t),\phi_\alpha(0,t)]}$ is unchanged by the time dependence and is still given by Eq.~\eqref{Eq:eq_LL_ecomphi}. In order to evaluate $\langle\phi_\alpha(x,t)\phi_\alpha(0,t)\rangle$, we use the time evolution of the $b_q$ operators from Eq.~\eqref{Eq:qu_bqTimeDep} in the definition of the bosonic fields 
    \begin{align}
        \phi(x,t)=-\sum_{q>0}\sqrt{\frac{2\pi}{qL}}e^{-\frac{q\eta}{2}}[e^{\imath\alpha q x}(w_q(t)b_{\alpha,q}+u_q(t)b^\dagger_{\alpha,-q})\notag \\+e^{-\imath\alpha qx}(w^*_q(t)b^\dagger_{\alpha,q}+u^*_q(t)b_{\alpha,-q})]\ .
    \end{align}
    Analogous to  
     the equilibrium case, we use $\langle b_q^\dagger b_q\rangle=f_b(q)$, to obtain \begin{align}
       & \langle\phi(x,t)\phi(x',t)\rangle = -\sum_{q>0}\frac{2\pi}{qL}e^{-q\eta}
        \notag \\
       & \times
       \Big\{e^{\imath\alpha q(x-x')}\left[(1-f_b(q)) |w_q(t)|^2 +f_b(q) |u_q(t)|^2\right]
        \notag\\
        &\ \ +e^{-\imath\alpha q(x-x')}\left[(1-f_b(q)) |u_q(t)|^2 +f_b(q) |w_q(t)|^2\right]\Big\} \ .
    \end{align}
    We again consider the zero temperature case with $f_b(q>0)=0$. This allows us to rewrite the desired exponential term from Eq.~\eqref{Eq:qu_LL_Calp} as follows 
    \begin{align}
        &-\frac{1}{2}\langle(\phi_\alpha(x,t)-\phi_\alpha(0,t))^2\rangle\notag\\
        &\phantom{--}=\sum_{q>0}\frac{2\pi}{qL}e^{-q\eta}\left(|w_q(t)|^2+|u_q(t)|^2\right)\notag\\ &\phantom{---=\,\sum_{q>0}}
        \times\left[-1+\frac{1}{2}e^{\imath\alpha qx}+\frac{1}{2}e^{-\imath\alpha qx}\right] \ .\label{Eq:qu_LL_phiphiexp}
    \end{align}
    Using $|w_q(t)|^2+|u_q(t)|^2=\cosh^2{(2\theta_q)}-\cos{(2v|q|t)}\sinh^2{(2\theta_q)}$, the above becomes 
    \begin{align}
       &-\frac{1}{2}\langle(\phi_\alpha(x,t)-\phi_\alpha(0,t))^2\rangle\notag\\
        &\phantom{--}=\sum_{q>0}\frac{2\pi}{qL}e^{-q\eta}\left(2\sin^2{(v|q|t)}\sinh^2{\left(2\theta_q\right)}+1\right)\notag\\ &\phantom{---=\,\sum_{q>0}}
        \times\left[-1+\frac{1}{2}e^{\imath\alpha qx}+\frac{1}{2}e^{-\imath\alpha qx}\right] \ .\label{Eq:qu_LL_phiphiexp2} 
    \end{align} 
    We define the momentum dependence of interaction parameter  in the quench  case as
    \begin{align}
        \sinh^2(2\theta_q) \approx \left(\frac{K-K^{-1}}{2}\right)^2\,e^{-\eps |q| } \equiv \gamma^2 \,e^{-\eps |q|} \ . \label{Eq:qu_LL_gamma}
    \end{align}
    The free term $\langle(\phi_\alpha(x,t)-\phi_\alpha(0,t))^2\rangle_{0}$ is equivalent to that in Eq.~\eqref{Eq:eq_phiphi_free}. 
    To compute the interaction term, we need to compute the $q$-sum, where we can use Eq.~\eqref{Eq:eq_LL_qsum} such that
    \begin{align}
        &\exp\left[{\kappa\sum_{q>0}\frac{2\pi}{L}e^{-\imath qz(x)}\sin^2{(v|q|t)}}\right]
        \notag\\ &\ \ 
        =\frac{[\cord{z(x)-2v|q|t}\cord{z+2v|q|t}]^{\kappa/4}}{\sin\left(\frac{\pi}{L}z(x)\right)^{\kappa/2}} \ .
    \end{align}
    The interaction term is then found to be 
    \begin{align}
       & e^{  -\frac{1}{2} \left[\langle(\phi_\alpha(x,t)-\phi_\alpha(0,t))^2\rangle
       -\langle(\phi_\alpha(x,t)-\phi_\alpha(0,t))^2\rangle_0
       \right]} \\
       &\phantom{--}
       =  \left|\frac{\cord{\imath(\eta+\epsilon)}}{\cord{\alpha x+\imath(\eta+\epsilon)}}\right|^{\gamma^2}    \notag \\
       &\phantom{--}
       \times \left|\frac{\cord{\alpha x+2vt+\imath(\eta+\epsilon)} }{\cord{2vt+\imath(\eta+\epsilon)} }\right|^{\gamma^2/2} \notag \\ 
       &\phantom{--}\times
       \left|\frac{ \cord{\alpha x-2vt+\imath(\eta+\epsilon)}}{ \cord{-2vt+\imath(\eta+\epsilon)}}\right|^{\gamma^2/2} \ .
       \label{Eq:qu_phiphi_int}
    \end{align} 
     Inserting the above and  Eq.~\eqref{Eq:phiphi_comm} into Eq.~\eqref{Eq:qu_LL_Calp}, 
     and taking the limit $\eta/L\to 0$,
     yields the correlation function for $\alpha$-movers. Adding together the right and left movers as in Eq.~\eqref{Eq:eq_LL_rhoTDlimit} gives the final expression
     \begin{align}
    &    \rho(x,t) = \rho_{1}^0(x,0)
        \left|\frac{\sin(\pi \imath \eps/L)}{\sin(\pi (x+\imath\eps)/L)}\right|^{\gamma^2}    \label{Eq:qu_LL_obdm}\\
       & \ \times   \left|\frac{\sin\left(\frac{\pi}{L}(x-2vt+\imath\epsilon)\right)\,\sin\left(\frac{\pi}{L}(x+2vt+\imath\epsilon)\right)}{\sin\left(\frac{\pi}{L}(-2vt+\imath\epsilon)\right)\,\sin\left(\frac{\pi}{L}(2vt+\imath\epsilon)\right)}\right|^{\gamma^2/2} \notag .
    \end{align}
	\begin{figure}[ t! ]
			\centering
			\includegraphics[width=8.6cm]{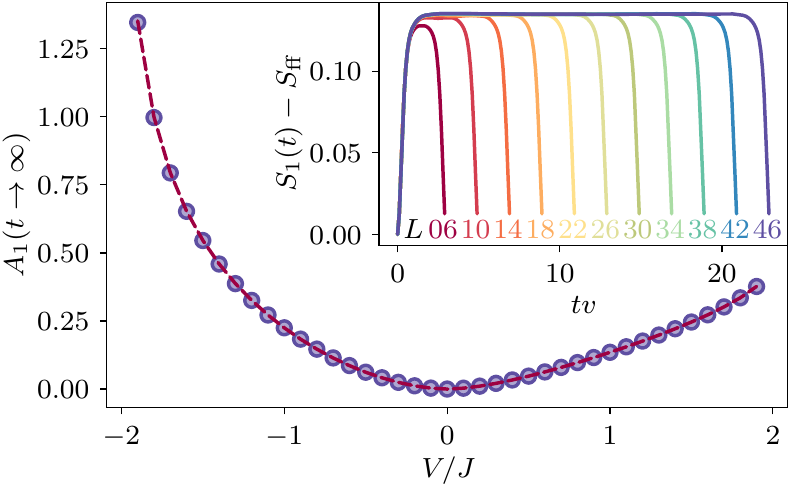}  
			\caption{\label{Fig:quench_timedependence_fss}Thermodynamic limit growth of the 1-particle von Neumann entanglement entropy as a function of the interaction strength $V/J$ after the quantum quench at $t=0$ obtained from the Luttinger liquid steady state limit (red, dashed line) in Eq.~\eqref{Eq:qu_LL_steadystate}. Blue circles show the result of finite size scaling of the plateau values shown in the inset. The inset depicts the first plateau of the entanglement entropy obtained by numerical integration from the time dependent one body density matrix in  Eq.~\eqref{Eq:qu_LL_obdm} for several system sizes $\Lb=2N$. We observe that the plateau size increases linearly with $L$ while the region where the entropy increases to the plateau and the time scale where it drops from the plateau is independent of the system length. In the thermodynamic limit, the average of the entropy and the plateau values agree with each other. The main panel demonstrates that the plateau value of the entropy coincides with the entropy obtained from the  steady state result for the 1-RDM in the thermodynamic limit.}
	\end{figure}
    The 1-RDM consists of the free part $ \rho_{1}^0(x,0)$ Eq.~\eqref{Eq:eq_LL_rho0TDlimit}, the interaction factor with exponent with $\gamma^2\neq\gamma^2_{\rm eq}$, and a time dependent oscillatory term with exponent $\gamma^2/2$. To obtain the one-particle entanglement entropy with Eq.~\eqref{Eq:eq_LL_Renyis}, we need to numerically compute the Fourier transform of Eq.~\eqref{Eq:qu_LL_obdm}. However, we can already extract information about the time dependence from the real space correlation function. We find that the entropy obtained from the LL correlation function is strictly periodic with period $\Delta t = L/(2v)$ and plateaus centered around $t_{n,\rm plateau} = L/(2v) (n+1/2), n\in\mathbb{N}_0 $ (see inset of Fig.~\ref{Fig:quench_timedependence_fss}) corresponding to times where all sine functions turn into cosine functions in Eq.~\eqref{Eq:qu_LL_obdm}. Because the size of the plateaus is proportional to $L$ and the time scale for increase and decrease from the plateaus is independent of $L$ (inset Fig.~\ref{Fig:quench_timedependence_fss}), the average converges to the plateau value in the thermodynamic limit. We compute the plateau values for many system lengths by numerically Fourier transforming Eq.~\eqref{Eq:qu_LL_obdm}, evaluating the entanglement entropy at $t_{0,\rm plateau}$, and performing finite size scaling to show the thermodynamic limit averaged entropies with blue circles in Fig.~\ref{Fig:quench_timedependence_fss}.  
    
    The steady state estimate of the entropy can also be analyzed by generalizing the scaling form introduced in Eq.~\eqref{eq:Adef} to include the post-quench waiting time:
    \begin{equation}
        A_\alpha(V/J,t) = \lim_{N\to\infty} S_\alpha(N,V/J,t) - \ln (N)\, .
    \end{equation}
    Its steady state value can be obtained from the 1-RDM in the $x/L\ll1$ limit (dashed, red line in Fig.~\ref{Fig:quench_timedependence_fss}) obtained from Eq.~\eqref{Eq:qu_LL_obdm}
    \begin{align}
        \rho_{t\rightarrow \infty}(x) &= \frac{\sin(k_F x)}{N\pi x} \left( \frac{\eps^2}{x^2 + \eps^2} \right)^{\gamma^2/2}  +\mathcal{O}\left(\frac{x}{L}\right)\ , \label{Eq:qu_LL_steadystate}
    \end{align}
    similar to the equilibrium case Eq.~\eqref{Eq:eq_LL_rhoTDlimit} and Eq.~\eqref{Eq:eq_LL_fqTDlimit} with $\gamma_{\rm eq}$ replaced by $\gamma$.

\subsection{Post-Quench Numerical Results}
\label{subsec:post_quench_numerics}

\subsubsection{Exact Diagonalization}
\label{ssec:post_quench_ed}

    % ED quench case
    We again use exact diagonalization to compute the waiting time dependence of the one-particle entanglement entropy after the quench. For this, we first obtain the ground state at $t<0$ for free fermions $\ket{\Psi(0)}$ and compute the time evolution using the full set of eigenstates $\ket{\Psi_\alpha}$ and eigenvalues $E_\alpha$ for the final Hamiltonian with interaction strength $V/J$ \cite{DelMaestro:2021ja}
    \begin{align}
       \ket{\Psi(t)} &= e^{-\imath t H}\ket{\Psi(0)} \nonumber \\
       &   = \sum_\alpha e^{-\imath E_\alpha t} \braket{\Psi_\alpha}{\Psi(0)} \ket{\Psi_\alpha} \ ,
    \end{align}
where we can exploit that $\braket{\Psi_\alpha}{\Psi(0)}$ is only non-zero for
$\ket{\Psi_\alpha}$ from the $q=0$ translational symmetry block. This still requires the full eigensystem of a dense block of the Hamiltonian whose size scales $\propto {2N\choose N}$ and a full diagonalization has a time complexity cubic in the Hamiltonian size, which limits us to a maximum of $N=13$ fermions on the lattice. 
   	\begin{figure}[ t! ]
			\centering
			\includegraphics[width=8.6cm]{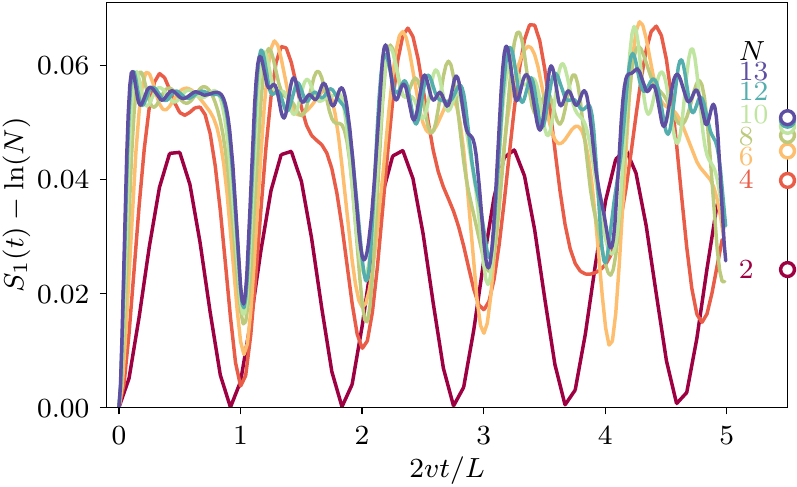}  
			\caption{\label{Fig:quench_timedependence}One-particle von Neumann entanglement entropy as a function of the rescaled waiting time $2vt/L$ after the quantum quench at $t=0$ for systems with $\Lb=2N$ sites at an interaction strength of $V/J=-0.5$. We observe a steep increase of the entropy on a very short time scale after the quench and then recurrences with length $N$ where the entropies oscillate around a constant steady state value (empty circles). The fast decrease in the distance between the steady state values with $N$ suggests fast convergence to the thermodynamic limit.}
	\end{figure}
%   
    % Description of time dependence of entropy
From $\ket{\Psi(t)}$, we obtain the density matrix ${\rho}(t) =
\ket{\Psi(t)}\bra{\Psi(t)}$ and trace out $N-1$ particle positions to
obtain $\rho_1(\abs{i-j},t)$ enabling computation of the 1-particle entanglement
entropy at each time $t$.  We indeed observe the recurrence time $\Delta t=
L/(2v) $ (Fig.~\ref{Fig:quench_timedependence}) predicted by the LL theory,
which indicates that after the quench, density waves propagate with
velocity $v$, (see Eq.~\eqref{Eq:qu_LL_vel}) through the lattice of length
$L$, where the maximal distance between two points is $L/2$ due to the
periodic boundary conditions. We show the waiting time dependence of the
von Neumann entropy for several lattice sizes (solid lines) in
Fig.~\ref{Fig:quench_timedependence} together with the steady state values
(empty circles) obtained by averaging the entropy $S_1-\ln(N)$ for times
after the initial increase. Even for these relatively small systems, the
fast decrease between consecutive steady state averages shows fast
convergence to the thermodynamic limit. Such advantageous finite size
scaling properties of the particle entanglement entropy were recently
reported \cite{DelMaestro:2021ja}.  To estimate errors in the steady state
averages, we use a blocking method \cite{Flyvbjerg.1989} by consecutively
averaging neighboring values in the time series and computing the error of the
mean in each averaging step until it reaches a plateau.  To further include errors due to the finite time step and the endpoint of the time series, we additionally divide the time series into the individual $N_b$ recurrence blocks with entropy averages $M_i$ and add the error of the means ${\rm mean}(M_i)/\sqrt{N_b}$, as well as the difference between the mean of the entropy time series and the average of the $M_i$ to the blocking error.

\subsubsection{Time Dependent Density Matrix Renormalization Group}
\label{ssec:post_quench_tdmrg}
   
    % DMRG after quench
To further enhance our ability to extrapolate to the thermodynamic limit
post-quench, we perform time evolution using approximate methods in
\texttt{ITensors.jl} \cite{itensor}.  To efficiently perform time evolution of the initial state obtained with DMRG as in the equilibrium case, we approximate the time evolution operator $e^{-\imath H \delta t}$ for a time step $\delta t$ by using a symmetrized second order Trotter decomposition \cite{Suzuki.1976,Paeckel.2019}  
    \begin{align}
       &e^{-\imath\delta t\, H} \approx e^{-\imath\delta t\, h_{1,2}/2} e^{-\imath\delta t\, h_{2,3}/2} \cdots   e^{-\imath\delta t\, h_{L,1}/2} \times  \notag\\
    &\;\;\times e^{-\imath\delta t\, h_{L,1}/2} e^{-\imath\delta t\, h_{L-1,L}/2}\cdots   e^{-\imath\delta t\, h_{1,2}/2} +\mathcal{O}(\delta t^3)\ , \label{Eq:qu_TrotterDecomp}
    \end{align} 
    where $h_{i,i+1}=-J   (c_{i+1}^\dagger c_i^{\phantom{\dagger}} + c_i^\dagger c_{i+1}^{\phantom{\dagger}}) + V   n_i n_{i+1}$. To derive Eq.~\eqref{Eq:qu_TrotterDecomp} the $J$-$V$ Hamiltonian 
    \begin{align}
        H &= \sum_{i=1}^L h_{i,i+1} =\sum_{i\, \rm even} h_{i,i+1} + \sum_{i\, \rm odd}h_{i,i+1} \equiv H_{\rm even} + H_{\rm odd}
    \end{align}
is split  into the two internally commuting parts $H_{\rm even}$ and
$H_{\rm odd}$. The commutator $[ H_{\rm even} ,H_{\rm odd}]$ is neglected,
which introduces an error $\mathcal{O}(\delta t^3)$. 
%To perform time evolution with a small error, 
To maintain accuracy, it is therefore
necessary to chose a small time step $\delta t$ such that performing time
evolution for a finite time interval $t$ can require a large number
$t/\delta t$ of time consuming applications of the operator. Only by using
GPUs for computing the time evolution were we able to perform the
calculation for systems up to $N=15$ fermions, which would be intractable with ED.   

\begin{figure}[ t! ]
			\centering
			\includegraphics[width=8.6cm]{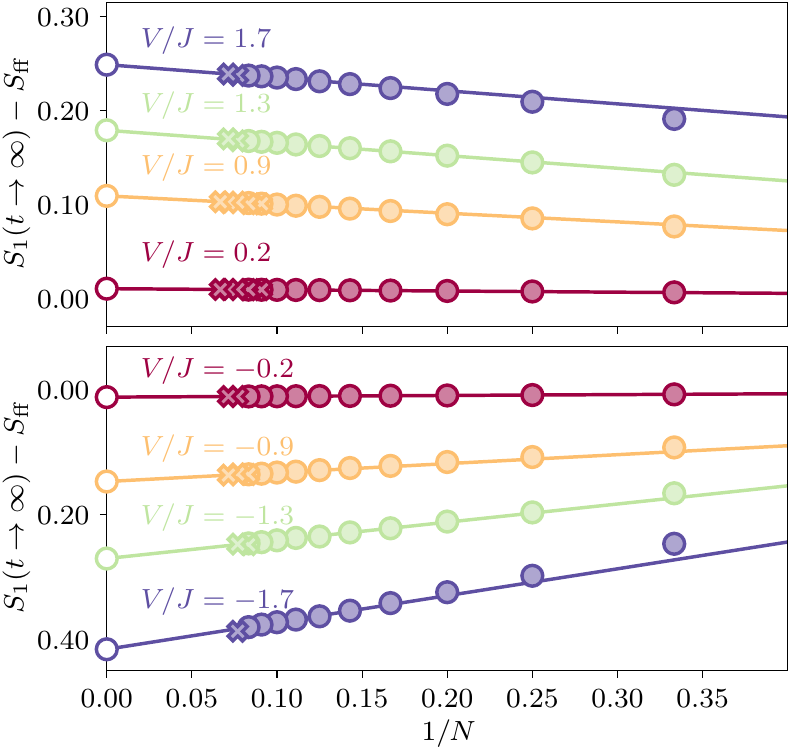}  
\caption{\label{Fig:quench_fsscaling}Finite size scaling of the steady state one-particle von Neumann entanglement entropy $S_1(t\rightarrow\infty)$ for various values of the post quench interaction strength $V/J$ obtained with ED (filled circles) and with DMRG (filled crosses). We estimate the thermodynamic limit values of the entropy (empty circles) using linear extrapolation to $1/N\rightarrow 0$.}
\end{figure}
    
\subsubsection{Finite size scaling}
\label{ssec:post_quench_fss}

We compute the 1-particle entanglement entropy for different interaction
strengths $V/J$ to again linearly extrapolate to the thermodynamic limit,
$1/N\rightarrow 0$ (Fig.~\ref{Fig:quench_fsscaling}).  In the case of small
interactions, we can perform the time evolution on V100 GPUs for systems with
up to $\Lb=30$ lattice sites, however, the ground states obtained with DMRG
require more memory for larger interactions $\gamma_{\rm eq}^2$ to perform
calculations to the same accuracy. While for intermediate interactions. $V/J\geq
1.3$ and $-0.9\leq V/J<0$, we achieve results up to $L=28$ sites, the
calculation exceeds GPU memory available to us for $L\geq28$ in the case of
very strong interactions $V/J \leq -1.3$.  Nonetheless, these additional points
of the 1-particle entanglement entropy obtained with the GPU accelerated tDMRG
allow for relative improvements of the thermodynamic limit extrapolation from finite size scaling by up to $1.2\,\%$ compared to using ED data alone.  We find that even for these relatively small systems, linear extrapolation accurately describes the data in the whole Luttinger liquid phase. For all parameters where we compute 1-particle entanglement entropies for both ED (circles) and DMRG (crosses), we find excellent agreement.

    \begin{figure}[ t! ]
			\centering
			\includegraphics[width=8.6cm]{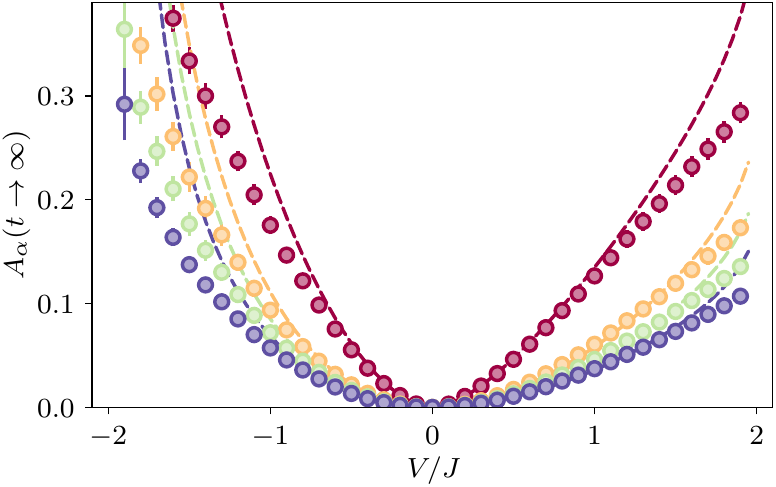}
\caption{\label{Fig:quench_tdlimit}1-particle entanglement entropies
extrapolated from the numerical steady state estimates to the thermodynamic
limit as a function of the post quench interaction strength $V/J$ (circles) for
\ren indices $\alpha=1$ (red), $\alpha = 2$ (yellow), $\alpha=3$ (green), and
$\alpha=5$ (blue). Dashed lines depict the corresponding theory predictions
from the steady states after the quench in the Luttinger liquid model using a
fixed interaction cutoff $\eps=0.84$ obtained from ground state calculations.
Similar to the equilibrium case, we find good agreement between LL prediction
and numerical data for moderate interaction strengths.}
	\end{figure}
    
    %Comparison with LL
\subsection{Comparison with Luttinger Liquid Theory}
\label{subsec:quench_LL_comparison}

Performing the finite size scaling for all computed interaction strengths $V/J$, we obtain the interaction dependence of the steady state 1-particle entanglement entropy in the thermodynamic limit (circles in Fig.~\ref{Fig:quench_tdlimit}) which we plot together with the entropy obtained from numerically computing the Fourier transform and numerically integrating the analytical steady state result from bosonization in Eq.~\eqref{Eq:qu_LL_steadystate} (dashed line, Fig.~\ref{Fig:quench_tdlimit}) for a fixed interaction cutoff $\eps=0.84$ determined in the ground state. We observe very similar agreement between results for the LL model and numerical results for the $J$-$V$ model as in the equilibrium case Fig.~\ref{Fig:eq_tdlimit_LL} when using the same cutoff.
	\begin{figure}[ t! ]
			\centering
			\includegraphics[width=8.6cm]{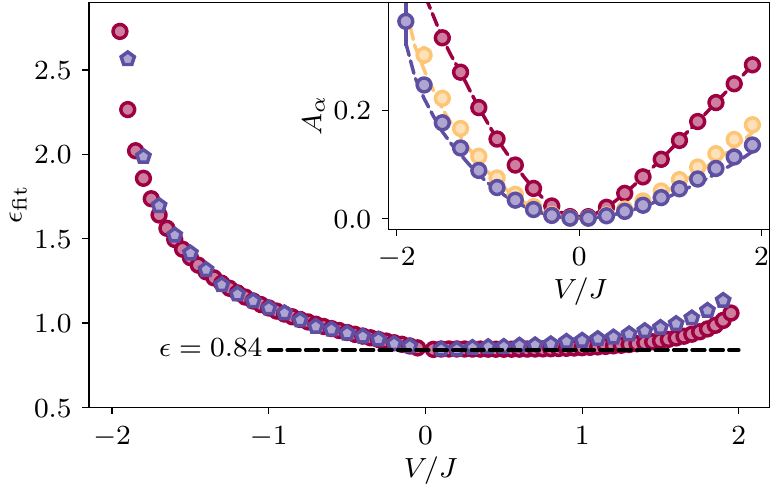}  
			\caption{\label{Fig:quench_fiteps}Interaction dependence of the effective cutoff $\eps_{\rm fit}$  (blue pentagons) obtained by fitting the steady state of the Luttinger liquid model at each interaction strength $V/J$ individually to the numerical data of the von Neumann entanglement entropy. For comparison, we show again the effective cutoff obtained from the ground state case (red circles) and find very good agreement with a quasi-plateau in the region $0<V/J<1$. The inset depicts numerical data for \ren entropies with $\alpha=1$ (red circles), $\alpha=2$ (yellow circles), and $\alpha=5$ (blue circles) together with the fitted field theory steady state predictions for the effective cutoff $\eps_{\rm fit}$.}
	\end{figure}
	
	%Fitting the cutoff
We again fit an effective interaction cutoff (blue pentagons in Fig.~\ref{Fig:quench_fiteps}) at each interaction strength $V/J$ separately to match the LL solution to the von Neumann entropy from the $J$-$V$ model (red circles in the inset of Fig.~\ref{Fig:quench_fiteps}). We find very good agreement with the interaction cutoff determined for the equilibrium ground state case (red circles in the main panel of Fig.~\ref{Fig:quench_fiteps}) which suggests that the parameter $\eps$ of the LL calculation can be fixed by numerical analysis of the $J$-$V$ model in the region $-0.5<J/V<1.5$, where the low energy LL approximation is most accurate resulting in $\eps=0.84$.

\section{Time Dependence of the spectrum of the post-quench 1-body reduced density matrix} 
\label{sec:time_dependence_rho}
%%%%%%%%%%%%%%%%
\begin{figure*}[ t! ]
\centering
\includegraphics[width=17.2cm]{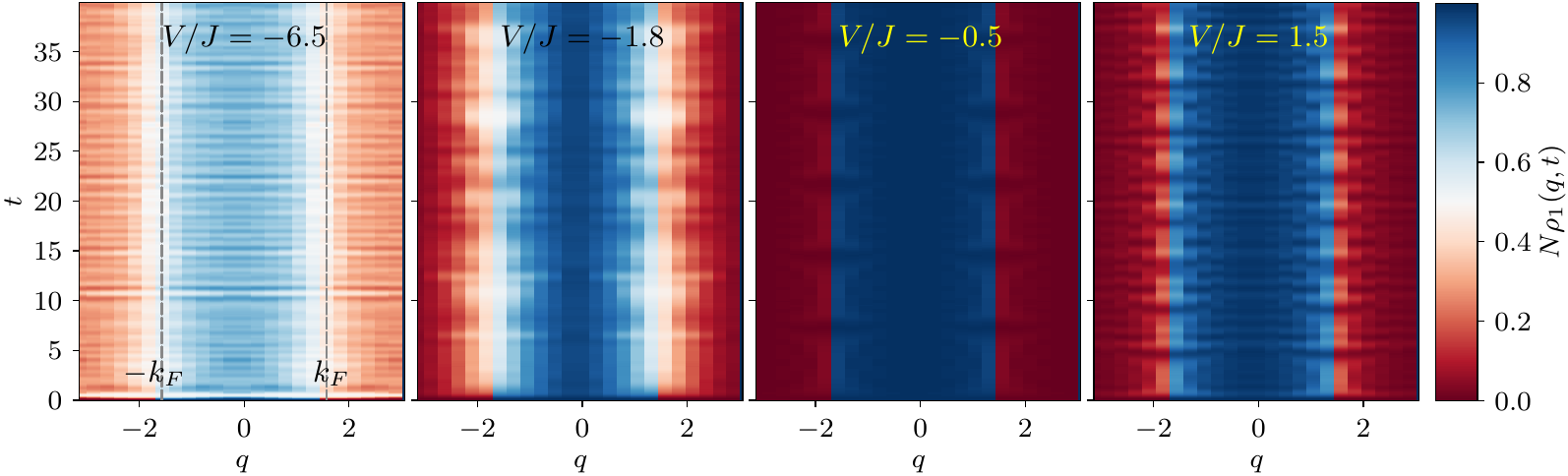}  
\caption{\label{Fig:obdm_eime_evolution}Exact diagonalization results for the time evolution of distribution function $N\rho_{1}(q,t)$ after a quantum quench for different values of the post quench interaction strength $V/J$.  The initial state at $t=0$ is the ground state of $N=12$ non-interacting fermions hopping on a ring of $\Lb=24$ sites. The dashed vertical lines mark the Fermi momenta, around which the fluctuations are more pronounced. Increasing the interaction strength increases the amplitude of fluctuations.}
\end{figure*}
%%%%%%%%%%%%%%%%

In this section, we utilize translational symmetry to monitor the time evolution of each eigenvalue of the 1-RDM. The initial state of free fermions $\ket{\Psi(0)}$ on the lattice is an eigenstate of the translation operator $T$, where $T\ket{\Psi(0)}=\ket{\Psi(0)}$. Also, the post-quench Hamiltonian commutes with the translation operator, $[H, T]=0$, ensuring that the time evolved state $\ket{\Psi(t)} = e^{-\imath t H}\ket{\Psi(0)}$ is an eigenstate of $T$ at all times, where $T\ket{\Psi(t)}=\ket{\Psi(t)}$.

If we now consider the elements of the two-point correlation matrix $\langle c_i^\dagger c_j^{\phantom{\dagger}}\rangle_{t}$ at time $t$ and use $c_i^\dagger c_j^{\phantom{\dagger}}=T^{\dagger} c_{i+1}^\dagger c_{j+1}^{\phantom{\dagger}} T^{\phantom{\dagger}}$ for $i,j\in \{1,\dots,\Lb-1\}$, we can write $\langle c_i^\dagger c_j^{\phantom{\dagger}}\rangle_{t}=\langle c_{i+1}^\dagger c_{j+1}^{\phantom{\dagger}}\rangle_{t}$. For elements that cross the boundary, we need to include the phase factor $(-1)^{N-1}$ due to the corresponding boundary conditions, \emph{e.g.},  $\langle c_i^\dagger c_{\Lb}^{\phantom{\dagger}}\rangle_{t}=(-1)^{N-1}\langle c_{i+1}^\dagger c_{1}^{\phantom{\dagger}}\rangle_{t}$. Therefore, the matrix is translationally invariant with a boundary phase $(-1)^{N-1}$ and can thus be diagonalized via Fourier transformation, where the diagonalized matrix represents the two-point correlation in terms of quasi-momenta modes, \emph{i.e.},
    %%%%%%%%%%%%%%%%
    \begin{align}
        \langle\Tilde{c}_q^\dagger\Tilde{c}_{q^\prime}^{\phantom{\dagger}}\rangle_{t}=\delta_{q,q^\prime}\langle n_q\rangle_{t}\ ,
    \end{align}
    %%%%%%%%%%%%%%%%
    where $\Tilde{c}_q=\Lb^{-1/2}\sum_j e^{-\imath qj}c_j$ and $n_q=\Tilde{c}^{\dagger}_q\Tilde{c}^{\phantom{\dagger}}_q$. Here, $q\in\{\left(2m-\Lb+b_N\right)\pi/\Lb: m=0,1,\dots,\Lb-1\}$, with $b_N=\frac{3-(-1)^N}{2}$.
    Accordingly, we obtain the momentum distribution function for the lattice fermions as 
     %%%%%%%%%%%%%%%%
    \begin{align}
        \rho_{1}(q,t)=\frac{1}{N}\langle n_q\rangle_{t}\ ,
    \end{align}
    %%%%%%%%%%%%%%%%
    where the canonical ensemble condition $\sum_q\langle n_q\rangle_{t}=N$ fixes the normalization of $\rho_{1}(q,t)$ such that $\sum_q \rho_{1}(q,t)=1$.
        Figure~\ref{Fig:obdm_eime_evolution} demonstrates the time evolution of $N\rho_{1}(q,t)$ for different interaction strengths obtained from exact diagonalization for a system with $N=12$ fermions on $\Lb=24$ lattice sites. At time $t=0$, we have the free fermionic occupation probabilities at zero temperature, where  $\langle n_q\rangle_{t=0}=1$ for $\vert q\vert\leq k_F$ and $\langle n_q\rangle_{t=0}=0$ otherwise. After the quench, the occupation probabilities start to change, and the quench in the LL phase ($-2<V/J<2$) generates fluctuations that are more visible near the Fermi level and increase with increasing interaction strength. This is consistent with the effective low energy LL description. For $V/J=-1.8$, the effective thermalization following the abrupt quantum quench starts to invoke the extremes of the energy spectrum, where the linear approximation of the spectrum no longer holds and band curvature effects may be important.  For comparison, we also consider a quench to an interaction strength of $V/J=-6.5$, which is outside of the Luttinger liquid phase.  As illustrated in Fig. \ref{Fig:obdm_eime_evolution}, the occupation probabilities show a flatter distribution and substantial fluctuations.

\begin{figure}[ t! ]
			\centering
			\includegraphics[width=8.6cm]{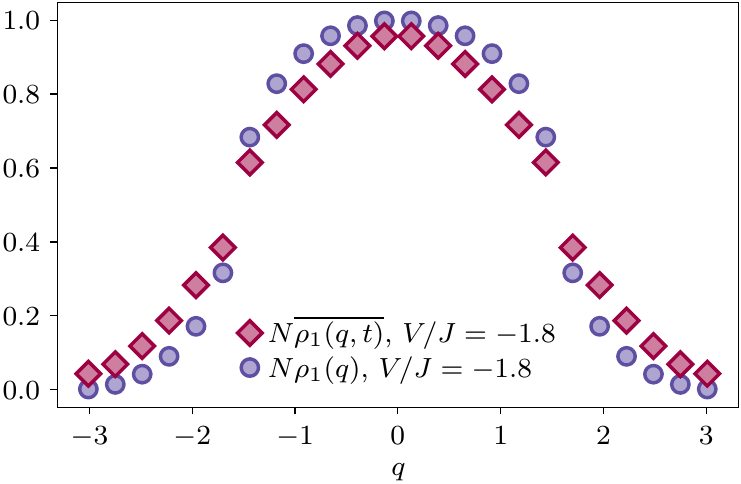}  
			\caption{\label{Fig:sp_obdmQavg}Comparison between the time average of the distribution function $N\overline{\rho_{1}(q,t)}$ (red diamonds) after a quench to interaction strength of $V/J=-1.8$ and the distribution function $N\rho_{1}(q)$ (blue circles) for the corresponding equilibrium case. The ED data is for a system of $N=12$ fermions at half-filling. }
	\end{figure}

    The time average of the distribution function $N\overline{\rho_{1}(q,t)}$ can provide information on quasi-thermalization after the quantum quench, as illustrated in Fig.~\ref{Fig:sp_obdmQavg}.
    %%%%%%%%%%%%%%%%
    %%%%%%%%%%%%%%%%
    Here, $\overline{\rho_{1}(q,t)}$ shows a wider distribution, \emph{i.e.}, larger entanglement entropy, if compared with the related equilibrium ground state distribution function $\rho_{1}(q)$. We can understand this by comparing the form of the steady-state 1-RDM $\rho_{t\rightarrow \infty}(x)$ (Eq.~\eqref{Eq:qu_LL_steadystate}) with the equilibrium one-body density function $\rho_{1}(x)$ (Eq.~\eqref{Eq:eq_LL_fqTDlimit}).  For the same interaction strength $V/J\neq0$, we have $\gamma>\gamma_{\rm eq}$, thus $\rho_{t\rightarrow \infty}(x)$ decays with $x$ faster than $\rho_{1}(x)$. Consequently, their Fourier transformation should exhibit the opposite behavior.

\begin{figure}[t!]
\centering
\includegraphics[width=8.6cm]{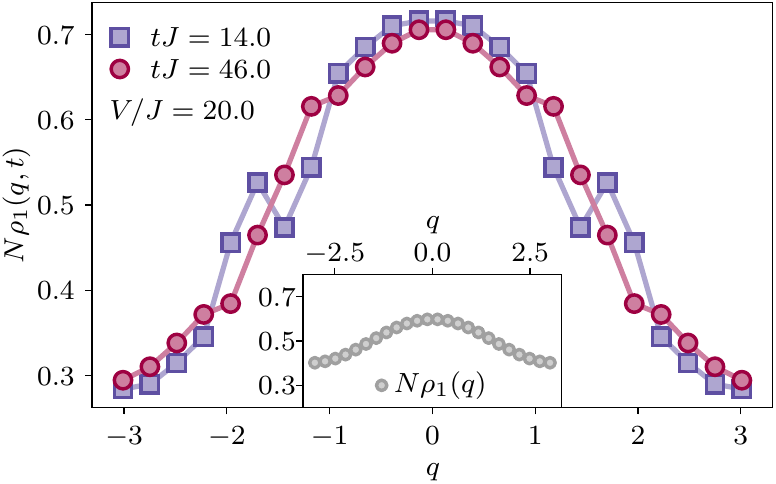}  
\caption{\label{Fig:sp_obdmQtdV20} Momentum distribution function at two different times, after a quantum quench at $t=0$ to a final interaction strength $V/J = 20$ deep in the density wave phase. The system consists of $N = 12$ fermions on a ring of $L = 24$ sites.  The inset shows the equilibrium ground state distribution function for the same interaction strength. Lines are a guide to the eye.}
    \end{figure}

    In Fig.~\ref{Fig:sp_obdmQtdV20} we show the momentum distribution for fixed waiting times $t$ after a strong interaction quench to $V/J = 20$, across the continuous phase transition to the density wave phase.  We observe similar behavior as discussed in the introduction in Fig.~\ref{Fig:sp_obdmQtd} for a quench across the discrete phase transition, where the momentum distribution can develop non-monotonic behavior as a function of $q$.
		 				
\section{Conclusion} 

In this paper we have reported on a comprehensive study of the one particle reduced density matrix and its associated von Neumann and \ren entanglement entropies in the $J$-$V$ lattice model of spinless fermions in one spatial dimension at half filling.  We have considered entanglement both in the ground state of the interacting model, as well as after an interaction quantum quench starting from an initial state of non-interacting fermions.  In both setups, we demonstrate that the 1-particle entanglement entropy is sensitive to the continuous and discrete phase transitions known to exist in this integrable model.

By carefully exploiting translation, reflection, and particle-hole symmetries of the lattice model in the presence of periodic boundary conditions, combined with advances in time-dependent density matrix renormalization group on massively parallel GPUs, we have pushed the boundaries of exact and approximate computations of the 1-particle entanglement entropy.  Specifically, we study system sizes up to $L=102$ sites in the ground state at half-filling, and up to $L=30$ after the quantum quench.  Here, periodic boundary conditions are essential to obtain the momentum distribution function via a simple Fourier transform of the one particle reduced density matrix. Access to these large system sizes are required for a reliable extrapolation to the thermodynamic limit.

For strong interaction quenches outside of the quantum liquid into the clustered solid ($V/J \ll -2$) or density wave ($V/J \gg 2$) phases, the momentum distribution function obtained from the spectrum of the one particle reduced density matrix can exhibit a non-monotonic dependence on momentum.  This behavior can occur for both small and large values of $q$, and may highlight dynamic signatures of the asymptotically flat momentum distributions in these two extreme limits of localized fermions.

For quenches within the quantum liquid regime, we can utilize continuum field theory calculations based on bosonization of the fermionic degrees of freedom within the Luttinger liquid phase ($\abs{V/J} \le 2$). With access to numerical predictions for $L\to\infty$, a  comparison between field theory and numerical results is possible. We use  a self-consistent approach for determining the interaction cutoff of the Luttinger model necessary due to the finite range nature of interactions in the lattice Hamiltonian. A fixed value of the cutoff is determined in the ground state, which can then be applied to the non-equilibrium post-quench dynamics. This provides a route to determining entanglement properties which depend on high energy degrees of freedom via bosonization.

Much work remains to be done to understand particle entanglement in interacting quantum many-body systems. For example, bosonization not only gives access to the one particle reduced density matrix, but higher order density matrices (e.g.\@ $n=2$) are also computable as correlation functions by similar methods.
More generally, the 1-particle reduced density matrix is the starting point for an expansion of the entanglement entropy in terms of higher order density matrices.  Such a research program will require generalizing the Kirkwood  expansion of the thermal entropy in terms of irreducible distribution functions \cite{Kirkwood42} to keep track of the required antisymmetrization of fermionic density matrices. 

\begin{acknowledgments} 
M.T. and B.R.\@ acknowledge funding by the Deutsche Forschungsgemeinschaft (DFG) under 
Grant No. 406116891 within the Research Training Group RTG 2522/1 and under 
grant RO 2247/11-1. This work was supported in part by the NSF under Grant No.~DMR-2041995. M.T.\@ acknowledges the University of Tennessee, Knoxville and the Institute of Advanced Materials and Manufacturing for hospitality during a research visit where a portion of this work was completed.
\end{acknowledgments}

\appendix

\section{Comparison with Lattice Green Function}
\label{App:latticeGF}
Consider a 1D lattice with $L$ sites and periodic boundary conditions, where lengths are measured in units of the lattice constant.  The Hamiltonian for $N$ free fermions is given by
\begin{equation}
H = -2 \sum_k \cos (k ) c_k^\dagger c_k \ ,
\end{equation}
where we measure energies in units of the hopping (i.e. $J=1$).  Here, our quantization condition for periodic boundary conditions is
\begin{equation}
k_m = \frac{2 \pi }{L}  m \quad \text{with} \quad m \in \left[-N,N\right) \ .
\end{equation}
For $N$ odd, the ground state is $| \Psi \rangle = \prod_{|k|<k_F} c^\dagger_k |0\rangle$ with $k_F=\pi(N-1)/L$ and the momentum distribution is 
\begin{equation}
n_k  = \langle c_k^\dagger c_k \rangle = 
\begin{cases}
1 & \quad |k| < k_F \\
0 & \quad \text{otherwise}
\end{cases} \ .
\end{equation}
To compute the Green function, we define the Fourier transform:
\begin{equation}
c_j = \frac{1}{\sqrt{\Lb}} \sum_k \mathrm{e}^{\imath k j } c_k \ .
\end{equation}
Thus we can write:
\begin{align}
\langle c_i^\dagger c_j \rangle &= \frac{1}{\Lb} \sum_{k,k'}  \mathrm{e}^{-\imath k i} \mathrm{e}^{\imath k' j }\langle c_k^\dagger c_{k'} \rangle \\ \nonumber
&= \frac{1}{\Lb} \sum_{|m| < (N-1)/{2}} \mathrm{e}^{-\imath 2 \pi m\,|i-j|/\Lb} \nonumber \\
&= \frac{1}{\Lb} \frac{\sin \frac{\pi N}{\Lb}|i-j|}{\sin \frac{\pi}{\Lb}|i-j|}\ .
\end{align}
The normalization condition that $\mathrm{Tr} \rho_1 = 1$ gives us:
\begin{equation}
\rho_1(|i-j|) = \frac{1}{N \Lb} \frac{\sin\left(\frac{\pi N}{L}|i-j|\right)}{\sin\left(\frac{\pi}{L}|i-j|\right)}\ .
\end{equation}
This gives the same result as Eq.~\eqref{Eq:eq_LL_rho0TDlimit} by using $x\rightarrow (i-j)$.  

%%%%%%%%%%%%%%%%    
\section{Definition of Symmetry Operators Based on Fermion Operators} \label{App:SymOperators}

    In this appendix, we provide additional details and a precise definition of a set of lattice symmetry operators, which are conserved by the $J$-$V$ Hamiltonian (Eq.~\eqref{Eq:eq_JVHam}). Starting from the definition of the operators by their action on the fermionic occupation basis, we write them in terms of the fermionic annihilation  $c_i^{\phantom{\dagger}}$ and creation  $c_i^\dagger$ operators, taking into account the anti-commutation relations $\{c_i^{\phantom{\dagger}},c_j^{\phantom{\dagger}}\}=0$, $\{c_i^{{\dagger}},c_j^{{\dagger}}\}=0$, and $\{c_i^{{\dagger}},c_j^{\phantom{\dagger}}\}=\delta_{i,j}$.
%%%%%%%%%%%%%%%%    
\subsection{Spatial inversion operator $R$} 
     We start by defining the spatial inversion operator $R=R^{-1}=R^{\dagger}$, which reflects the fermionic occupation numbers across the center of the lattice, \emph{e.g.}, $R\ket{011001}=\ket{100110}$, where $0$ and $1$ denote the empty and occupied sites respectively. If we define the occupation basis in terms of the action of the creation operator $c_j^{\dagger}$ on the vacuum state $\ket{0}$, then we have, for example, $\ket{011001}=c_2^{\dagger}c_3^{\dagger}c_6^{\dagger}\ket{0}$, with the convention of having the site labels $j$ of $c_j^{\dagger}$ in an ascending order. For the above case, we can write $Rc_2^{\dagger}c_3^{\dagger}c_6^{\dagger}\ket{0}=c_1^{\dagger}c_4^{\dagger}c_5^{\dagger}\ket{0}$, where $R\ket{0}=\ket{0}$.
 
    In general, for a lattice with $\Lb$ sites, $R$ sets the occupation state of site $j$ in the resulting basis ket to the occupation state of site $\Lb-j+1$ in the original basis ket. Based on this, if we define the operator $R^\prime$ such that
    %%%%%%%%%%%%%%%%
    \begin{align}
    R^\prime c_j^{\dagger}=c_{\Lb-j+1}^{\dagger}R^\prime\ ,
    \end{align}
    %%%%%%%%%%%%%%%%
    with $R^\prime\ket{0} =\ket{0}$ and then apply $R^\prime$ on the above example, we find
    %%%%%%%%%%%%%%%%
    \begin{align}
    R^\prime\ket{011001}&=R^\prime c_2^{\dagger}c_3^{\dagger}c_6^{\dagger}\ket{0}=c_5^{\dagger}c_4^{\dagger}c_1^{\dagger}\ket{0} \notag\\ &=-c_1^{\dagger}c_4^{\dagger}c_5^{\dagger}\ket{0}=-\ket{100110} \ .
    \end{align}
    %%%%%%%%%%%%%%%%
    Therefore, $ R^\prime \not\equiv R$, due to the appearance of the negative phase factor, where, in general, this phase factor depends on the number of fermions $N$ described by the occupation basis ket and it is given by $(-1)^{N(N-1)/2}$.
 
    To obtain a proper definition of $R$, we consider attaching the fermionic strings  $K_j^{\phantom{\dagger}}=e^{\sum_{k=1}^{j-1}-\imath\pi{n}_k}$ and $K_j^{\dagger}=K_j^{-1}=e^{\sum_{k=1}^{j-1}\imath\pi{n}_k}$ to the operators $c_j^{\phantom{\dagger}}$ and $c_j^{\dagger}$, respectively, where ${n}_j=c_j^{\dagger}c_j^{\phantom{\dagger}}$. Having the relations $K_j^{\dagger}\ket{0}=\ket{0}$ and for $j\leq i$ $\left[K_j^{\dagger},c_i^{\dagger}\right]=0$, allows us to insert the fermionic strings in the expression of any basis ket, for example, $c_2^{\dagger}c_3^{\dagger}c_6^{\dagger}\ket{0}=c_2^{\dagger}c_3^{\dagger}c_6^{\dagger}K_2^{\dagger}K_3^{\dagger}K_6^{\dagger}\ket{0}=c_2^{\dagger}K_2^{\dagger}c_3^{\dagger}K_3^{\dagger}c_6^{\dagger}K_6^{\dagger}\ket{0}$. Also, we have the commutation relations $\left[K_i^{\phantom{\dagger}}c_i^{\phantom{\dagger}},K_j^{\phantom{\dagger}}c_j^{\phantom{\dagger}}\right]=0$, $\left[c_i^{\dagger}K_i^{\dagger},c_j^{\dagger}K_j^{\dagger}\right]=0$, and for $i\neq j$, $\left[c_i^{\dagger}K_i^{\dagger},K_j^{\phantom{\dagger}}c_j^{\phantom{\dagger}}\right]=0$. We take advantage of the above commutation relations and define:
    %%%%%%%%%%%%%%%%
    \begin{align}
    R c_j^{\dagger}K_j^{\dagger}=c_{\Lb-j+1}^{\dagger}K_{\Lb-j+1}^{\dagger}R\ ,\label{Eq:Rdef}
    \end{align}
    %%%%%%%%%%%%%%%%
    with $R\ket{0} =\ket{0}$. Taking the Hermitian conjugate of the above equation and using $R^2=1$ , yields. $R K_j^{\phantom{\dagger}}c_j^{\phantom{\dagger}} =K_{\Lb-j+1}^{\phantom{\dagger}}c_{\Lb-j+1}^{\phantom{\dagger}}R$. If we now use $R$ instead of $R^\prime$ in the previous example, we get 
    %%%%%%%%%%%%%%%%
    \begin{align}
    R\ket{011001}&=Rc_2^{\dagger}K_2^{\dagger}c_3^{\dagger}K_3^{\dagger}c_6^{\dagger}K_6^{\dagger}\ket{0} \notag\\ &=c_5^{\dagger}K_5^{\dagger}c_4^{\dagger}K_4^{\dagger}c_1^{\dagger}K_1^{\dagger}\ket{0} \notag\\ &=c_1^{\dagger}K_1^{\dagger}c_4^{\dagger}K_4^{\dagger}c_5^{\dagger}K_5^{\dagger}\ket{0}\notag\\
    &=c_1^{\dagger}c_4^{\dagger}c_5^{\dagger}\ket{0}=\ket{100110} \ ,
    \end{align}
    %%%%%%%%%%%%%%%%
    where we reordered the commuting operators $c_j^{\dagger}K_j^{\dagger}$ after the action of $R$ takes place, then we removed the fermionic strings $K_j^{\dagger}$, similarly to their insertion. Accordingly, defining $R$ as in Eq.~\eqref{Eq:Rdef} prevents the appearance of any negative factors. 
   
   To simplify the definition in Eq.~\eqref{Eq:Rdef}, we first consider the action of $R$ on the occupation number operators ${n}_j=c_j^{\dagger}c_j^{\phantom{\dagger}}$, which is
    %%%%%%%%%%%%%%%%
    \begin{align}
    R {n}_j={n}_{\Lb-j+1}R\ .\label{Eq:Rnj}
    \end{align}
    %%%%%%%%%%%%%%%%
    Hence, $RK_j^{\phantom{\dagger}}=Re^{\sum_{k=1}^{j-1}-\imath\pi{n}_k}=e^{\sum_{k=\Lb-j+2}^{\Lb}-\imath\pi{n}_k}R$ and thus, $R c_j^{\dagger}=c_{\Lb-j+1}^{\dagger}K_{\Lb-j+1}^{\dagger}RK_j^{\phantom{\dagger}}=c_{\Lb-j+1}^{\dagger}e^{\imath\pi{n}_{\Lb-j+1}}e^{-\imath\pi\hat{N}}R$, where $\hat{N}=\sum_{k=1}^{\Lb}{n}_k$. Using $c_j^{\dagger}{n}_j=0$, we finally arrive at the useful results: 
    %%%%%%%%%%%%%%%%
    \begin{equation}
    \begin{aligned}
    R c_j^{\dagger}&=c_{\Lb-j+1}^{\dagger}e^{-\imath\pi\hat{N}}R \\
    Rc_j^{\phantom{\dagger}}&=e^{\imath\pi\hat{N}}c_{\Lb-j+1}^{\phantom{\dagger}}R\ .
    \end{aligned}
    \label{Eq:RdefR}
    \end{equation}
    %%%%%%%%%%%%%%%%
%%%%%%%%%%%%%%%%    
\subsection{Particle-Hole exchange operator $P$} 
        The particle-hole exchange operator $P$ changes the occupation states of each site on a basis ket of spinless fermions by emptying the occupied sites and occupying the empty ones, \emph{e.g.}, $P\ket{010011}=\ket{101100}$, where $P=P^{-1}=P^{\dagger}$. 
        
        Similar to spatial inversion operator case, to avoid negative phase factors, we include the fermions strings in the definition of $P$ as 
    %%%%%%%%%%%%%%%%
    \begin{align}
    P c_j^{\dagger}K_j^{\dagger}=K_j^{\phantom{\dagger}}c_j^{\phantom{\dagger}}P\ ,\label{Eq:Pdef}
    \end{align}
    %%%%%%%%%%%%%%%%
    and $P\ket{0}=\ket{11\dots1}$. Consequently, 
    %%%%%%%%%%%%%%%%
    \begin{align}
    P {n}_j=\left(1-{n}_j\right)P\ ,\label{Eq:Pnj}
    \end{align}
    %%%%%%%%%%%%%%%%
    and $PK_j^{\phantom{\dagger}}=Pe^{\sum_{k=1}^{j-1}-\imath\pi{n}_k}=(-1)^{j-1}K_j^{\dagger}P$. Thus, we simplify  Eq.~\eqref{Eq:Pdef} and obtain
    %%%%%%%%%%%%%%%%
    \begin{align}
    P c_j^{\dagger}=(-1)^{j-1}c_j^{\phantom{\dagger}}P\ .\label{Eq:PdefR}
    \end{align}
    %%%%%%%%%%%%%%%%    
%%%%%%%%%%%%%%%%    
\subsection{Translation operator $T$}
    We now consider translations, where the unitary operator $T$ rotates the occupation basis of the fermionic ring by one site, \emph{e.g.}, $T\ket{010011}=\ket{101001}$, where $T^{\Lb}=1$ and $T\ket{0}=\ket{0}$. Similarly to the previous operators, we define $T$ as
    %%%%%%%%%%%%%%%%    
    \begin{equation}
    \begin{aligned}
    T c_j^{\dagger}K_j^{\dagger}&=c_{j+1}^{\dagger}K_{j+1}^{\dagger}T \\
    T c_{\Lb}^{\dagger}K_{\Lb}^{\dagger}&=c_1^{\dagger}K_1^{\dagger}T\ ,
    \end{aligned}
    \label{Eq:Tdef}
    \end{equation}
    %%%%%%%%%%%%%%%% 
    hence
    %%%%%%%%%%%%%%%%
    \begin{equation}
    \begin{aligned}
    T {n}_j&={n}_{j+1}T \\
    T {n}_{\Lb}&={n}_1T\ .
    \end{aligned}
    \label{Eq:Tnj}
    \end{equation}
    %%%%%%%%%%%%%%%%
    To simplify the definition in Eq.~\eqref{Eq:Tdef}, we use Eq.~\eqref{Eq:Tnj} to write $K_{j+1}^{\dagger}TK_{j}^{\phantom{\dagger}}=e^{\sum_{k=1}^{j}\imath\pi{n}_k}Te^{\sum_{k=1}^{j-1}-\imath\pi{n}_k}=e^{\imath\pi{n}_1}T$ and $K_{1}^{\dagger}TK_{\Lb}^{\phantom{\dagger}}=Te^{\sum_{k=1}^{\Lb-1}-\imath\pi{n}_k}=e^{\imath\pi{n}_1}e^{-\imath\pi{N}}T$, resulting in
    %%%%%%%%%%%%%%%%    
    \begin{equation}
    \begin{aligned}
    T c_j^{\dagger}&=c_{j+1}^{\dagger}e^{\imath\pi{n}_1}T \\
    T c_{\Lb}^{\dagger}&=c_1^{\dagger}e^{-\imath\pi\hat{N}}T\ ,
    \end{aligned}
    \label{Eq:Tdef2}
    \end{equation}
    %%%%%%%%%%%%%%%% 
    where we used $c_1^{\dagger}e^{\imath\pi{n}_1}=c_1^{\dagger}$.

%\bibliography{bibliography}

\end{document}